\documentclass[11pt,a4paper]{article}
\pdfoutput=1
\usepackage{jheppub}
\usepackage{booktabs}
\usepackage{multirow}
\usepackage{graphicx}
\usepackage{epsfig}
\usepackage{lscape}
\usepackage{rotating}
\usepackage{hyperref}
\usepackage{amsmath}
\usepackage{lineno}
\usepackage{hyperref}

\def\micro{{\tt micrOMEGAs}}
\def\nmssmtools{{\tt NMSSMTools}}
\def\smodels{{\tt SModelS}}
\def\ma5{{\tt MadAnalysis5}}
\def\sushi{{\tt SusHi}}
\newcommand{\gtap}{\stackrel{\displaystyle >}{\,_{\! \,_{\displaystyle
\sim}}}} 
\newcommand{\ltap}{\stackrel{\displaystyle <}{\,_{\! \,_{\displaystyle
\sim}}}}
\def\beq{\begin{equation}}
\def\eeq{\end{equation}}
\def\bea{\begin{eqnarray}}
\def\eea{\end{eqnarray}}
\def\ba{\begin{array}}
\def\ea{\end{array}}
\def\nn{\nonumber}
\def\l{\lambda}
\def\k{\kappa}
\def\tb{\tan\beta}
\def\b{\beta}
\def\ds{\displaystyle}

\title{Status and prospects of the nMSSM after LHC Run-1}
\author[a]{D. Barducci}
\author[a]{G. B\'elanger}
\author[b]{C. Hugonie}
\author[c]{A. Pukhov}

\affiliation[a]{LAPTh, Universit\'e de Savoie Mont Blanc, CNRS, B.P.110, F-74941 Annecy-le-Vieux, France}
\affiliation[b]{LUPM, UMR 5299, CNRS, Universit\'e de Montpellier, 34095 Montpellier, France}
\affiliation[c]{Skobeltsyn Inst. of Nuclear Physics, Moscow State Univ., Moscow 119992, Russia}

\emailAdd{barducci@lapth.cnrs.fr}
\emailAdd{belanger@lapth.cnrs.fr}
\emailAdd{cyril.hugonie@umontpellier.fr}
\emailAdd{pukhov@lapth.cnrs.fr}

\abstract{The new minimal supersymmetric standard model (nMSSM), a variant of the general next to minimal supersymmetric standard model (NMSSM) without $Z_3$ symmetry, features a naturally light singlino with a mass below 75~GeV. In light of the new constraints from LHC Run-1 on the Higgs couplings, sparticles searches and flavour observables, we define the parameter space of the model which is compatible with both collider and dark matter (DM) properties. Among the regions compatible with these constraints, implemented through {\tt NMSSMTools}, {\tt SModelS} and \ma5, only one with a singlino lightest supersymmetric particle (LSP) with a mass around 5~GeV can explain all the DM  abundance of the universe, while heavier mixed singlinos can only form one of the DM components. Typical collider signatures for each region of the parameter space are investigated. In particular, the decay of the 125~GeV Higgs into light scalars and/or pseudoscalars and the decay of the heavy Higgs into charginos and neutralinos, provide distinctive signatures of the model. Moreover, the sfermion decays usually proceed through heavier neutralinos rather than directly into the LSP, as the couplings to the singlino are suppressed. We also show that direct detection searches are complementary to collider ones, and that a future ton-scale detector could completely probe the region of parameter space with a LSP mass around 65~GeV. }

\begin{document}

\date\today

 \begin{flushright}
 \hspace{3cm} LAPTH-053/15\\
 \hspace{3cm} LUPM:15-015
 \end{flushright}

\maketitle

\section{Introduction}
The discovery of a Higgs boson with a mass of 125~GeV at the Large Hadron Collider (LHC)~\cite{Aad:2012tfa,Chatrchyan:2012ufa} can be viewed as an argument in favour of supersymmetry (SUSY) since a light Higgs boson is a landmark of this theory. However the mass of the new particle is only within a few GeV of the maximum value predicted in the minimal supersymmetric standard model (MSSM) and requires large contributions from the stop sector, thus raising the issue of fine-tuning~\cite{Barbieri:1987fn,Hall:2011aa}. In the next-to-minimal supersymmetric extension of the standard model, the NMSSM~\footnote{For a review, see refs.~\cite{Maniatis:2009re,Ellwanger:2009dp}.}, the fine-tuning issue is not as severe because of additional contributions to the lightest Higgs doublet mass, derived from the extra singlet superfield~\cite{Ellwanger:2011mu,Cao:2012fz,Ellwanger:2012ke,Perelstein:2012qg,Agashe:2012zq,Gherghetta:2012gb,Cheng:2013fma,Kim:2013uxa,Fowlie:2014faa}. The NMSSM has the nice additional feature that the $\mu$ term is generated from the vacuum expectation value (VEV) of the new singlet field and is thus naturally at the SUSY scale, therefore solving the so-called $\mu$-problem~\cite{Kim:1983dt}. For these reasons the discovery of the Higgs at the LHC has triggered a renewed interest in the NMSSM and phenomenological studies abound~\cite{Munir:2013wka,Belanger:2014roa,Ellwanger:2014hia,Jeong:2014xaa,King:2014xwa,Ellwanger:2014hca,Bomark:2015fga,Chakraborty:2015xia,Potter:2015wsa}. The main focus has been on the Higgs sector, since the extra singlet can lead to new collider signatures, in particular when light, as the Standard Model (SM) like Higgs state $h$ with $m_h \sim 125$~GeV can decay into light singlet like scalars or pseudoscalars. Moreover its spin 1/2 SUSY partner, the singlino, can be at a mass well below the electroweak (EW) scale, giving also rise to peculiar SUSY signatures, especially when it is the lightest SUSY particle (LSP) and very light~\cite{Ellwanger:2014hia,Ellwanger:2014hca,Potter:2015wsa,Das:2012rr}. The NMSSM also provides a dark matter (DM) candidate, the LSP neutralino, and its properties have been analysed, both in the NMSSM with arbitrary parameters at the SUSY scale~\cite{Belanger:2005kh, Vasquez:2012hn} and in the grand unified theory (GUT) scale constrained models~\cite{Hugonie:2007vd,Belanger:2008nt,Ellwanger:2014dfa}. In general the predictions are similar to those of the MSSM, but special features including the possibility of a light neutralino LSP with an important singlino component, can increase the annihilation into Higgs final states or resonant annihilation through a singlet Higgs. Light neutralinos LSP can also escape astrophysical constraints~\cite{Vasquez:2010ru,AlbornozVasquez:2011js,AlbornozVasquez:2012px}. 

These studies were conducted within the framework of the $Z_3$ invariant NMSSM. However the superpotential of the general NMSSM does not necessarily possess this accidental $Z_3$ symmetry. The new minimal supersymmetric model (nMSSM), sometimes also called minimal next-to-minimal supersymmetric model (MNSSM)~\cite{Panagiotakopoulos:1999ah,Panagiotakopoulos:2000wp,Dedes:2000jp}, features instead a global discrete R-symmetry which forbids the singlet cubic self interaction in the superpotential, while avoiding problems with domain walls due to the $Z_3$ symmetry. Although the field content is the same as that of the ($Z_3$ invariant) NMSSM, the different superpotential and soft SUSY breaking terms lead to a peculiar phenomenology~\cite{Panagiotakopoulos:2001zy,Menon:2004wv,Barger:2005hb,Barger:2006dh,Huber:2006wf,Barger:2006kt,Barger:2006sk,Barger:2007nv,Balazs:2007pf,Ham:2007wu,Huber:2007vva,Chun:2008pg,Ham:2008cg,Cao:2009ad,Cao:2010fi,Ishikawa:2014owa}. The most striking feature of the model is that there is no mass term for the pure singlino. Only mixing effects with higgsinos can raise the singlino mass to the EW scale (up to $\sim75$~GeV~\cite{Hesselbach:2007te}). The singlino is therefore naturally light, and the LSP generally contains a large singlino component, thus guaranteeing a phenomenology rather different from that of the MSSM both at colliders and in DM searches. Yet, all the phenomenological studies of the nMSSM have overlooked the possibility of a very light singlino LSP (below 5~GeV) in agreement with DM constraints. In addition, the results from LHC Run-1 and the prospects for Run-2 are still unexplored in this model. This is the gap we intend to fill here.

In this paper we explore the parameter space of the nMSSM with unified conditions at the GUT scale, that is compatible with the latest Higgs results, with LHC searches for SUSY particles in Run-1 and with DM constraints. For this we rely on \nmssmtools~\cite{Ellwanger:2005dv,Ellwanger:2006rn} for the calculation of the spectrum and constraints on the Higgs sector, on \smodels~\cite{Kraml:2014sna,Kraml:2013mwa} for comparisons with limits on simplified models, on \ma5~\cite{Conte:2012fm,Conte:2014zja,Dumont:2014tja} for a more complete implementation of LHC searches for sparticles and on \micro~\cite{Belanger:2006is,Belanger:2008sj,Belanger:2014vza} for the computation of the DM observables including relic density, direct detection (DD) and indirect detection (ID).

We show that this combined set of requirements lead to strong constraints on the model. We found that the allowed regions have very specific characteristics: they contain either a very light singlino LSP below 5~GeV, a mixed singlino-higgsino LSP with a mass around 45~GeV or 65~GeV, or a bino LSP with a mass around 65~GeV. We have also checked that the allowed regions in parameter space of the general model with arbitrary parameters at the SUSY scale ({\it i.e.} without unified conditions at the GUT scale) share the same general characteristics. After having discussed the main constraints on the model, we analyse for each region some distinctive signatures that can arise at the LHC Run-2. Here we consider both searches for new Higgs states and sparticles. The complementarity between collider and DD searches is also highlighted. Moreover, we investigate the potential of ID to probe this model, which turns out to be quite limited, except for very peculiar kinematics which could lead to an enhanced gamma-ray line~\cite{Chalons:2011ia,Das:2012ys,Chalons:2012xf}. 

The paper is organized as follows. Section~\ref{sec:sec2} describes the model. Section~\ref{sec:sec3} contains the results of the parameter scan and the description of the different allowed regions. The LHC phenomenology of each region and the complementarity with DM searches is explored in detail in Section~\ref{sec:sec4}. Finally benchmark points corresponding to the channels with interesting signatures at the LHC are provided. Section~\ref{sec:sec5} contains our conclusions.

\section{The nMSSM}
\label{sec:sec2}

The MSSM is defined by promoting each SM field $\Phi$ into a superfield $\widehat{\Phi}$, doubling the Higgs fields with two $SU(2)_L$ doublets $H_u$, $H_d$ and imposing R-parity conservation. SUSY breaking is assumed to occur in an invisible sector and to be mediated through gravitational interactions to the visible sector. The resulting theory contains a number of soft SUSY breaking terms proportional to powers of the SUSY breaking scale $M_{\rm susy}$. Unfortunately, a realistic realisation of EW symmetry breaking in the MSSM requires the presence of the so called $\mu$-term in the superpotential, coupling directly the two Higgs fields $H_u$ and $H_d$:
\beq
W_{\rm MSSM} = \mu \widehat{H}_u \widehat{H}_d + h_u\widehat{Q}\widehat{U}^c\widehat{H}_u + h_d\widehat{Q}\widehat{D}^c\widehat{H}_d + h_e\widehat{L}\widehat{E}^c\widehat{H}_d \; ,
\eeq
with values of the arbitrary $\mu$ parameter close to $M_{\rm susy}$. There exist explanations for such a value of the $\mu$-term, alas, all in extended settings~\cite{Giudice:1988yz}. The easiest solution to the $\mu$-problem is to introduce an extra gauge singlet S, coupled to the Higgs doublets and whose VEV is naturally of the order of $M_{\rm susy}$. This leads to the simplest extension of the MSSM, the NMSSM with a cubic (renormalisable) superpotential
\beq\label{eq:WNMSSM}
W_{\rm NMSSM} = \l \widehat{S}\widehat{H}_u\widehat{H}_d + \frac{1}{3}\k\widehat{S}^3+ h_u\widehat{Q}\widehat{U}^c\widehat{H}_u + h_d\widehat{Q}\widehat{D}^c\widehat{H}_d + h_e\widehat{L}\widehat{E}^c\widehat{H}_d \; .
\eeq
The presence of the singlet cubic self interaction is necessary in order to avoid a global $U(1)$ Peccei-Quinn (PQ) symmetry which would lead to massless singlet states, experimentally excluded~\cite{Ellwanger:2009dp}. The corresponding soft SUSY breaking potential is given by
\bea\label{eq:VNMSSM}
V_\mathrm{NMSSM} &=&
m_{H_u}^2 | H_u |^2 + m_{H_d}^2 | H_d |^2 + m_{S}^2 | S |^2 + \left( \l A_\l H_u H_d S + \frac{1}{3} \k A_\k S^3 + \mathrm{h.c.}\right) \nn \\
&& +\, m_Q^2|Q^2| + m_U^2|U_R^2| + m_D^2|D_R^2| + m_L^2|L^2| + m_E^2|E_R^2| \nn \\
&& + \left( h_u A_u Q H_u U_R^c - h_d A_d Q H_d D_R^c - h_e A_e L H_d E_R^c + \mathrm{h.c.}\right) \nn \\
&& +\, M_1\, \widetilde{B}\, \widetilde{B} \, + \, M_2\, \widetilde{W}\, \widetilde{W} \, + \, M_3\, \widetilde{g}\, \widetilde{g} \; .
\eea

The standard NMMSM given by Eqs.~(\ref{eq:WNMSSM}-\ref{eq:VNMSSM}) possesses a global $Z_3$ symmetry under which all superfields are multiplied by $e^{2i\pi/3}$. Once this discrete symmetry is spontaneously broken during the phase transition associated with the EW symmetry breaking in the early universe,  cosmologically dangerous domain walls are generated ~\cite{Vilenkin:1984ib}. It has been argued in ~\cite{Abel:1995wk,Abel:1995uc} that these walls become harmless if they disappear before nucleosynthesis, which requires the presence in the effective potential of $Z_3$ breaking terms of magnitude
\beq
\delta V \sim O(1\ {\rm MeV})^4 \, .
\eeq
The magnitude of the $Z_3$ breaking terms corresponds to the presence in the superpotential or in the K\"{a}hler potential of $Z_3$ breaking operators suppressed by one inverse power of the Planck mass, $M_{\rm Planck}$. However, these $Z_3$ breaking (non-renormalisable) terms involving the singlet S induce divergent tadpoles~\cite{Nilles:1982mp,Lahanas:1982bk,Ellwanger:1983mg,Nilles:1997me,Bagger:1993ji,Jain:1994tk,Bagger:1995ay} of the form
\beq
\delta W = \Xi_F\, M_{\rm susy}\, M_{\rm Planck}\, S \; , \qquad \delta V = \Xi_S\, M_{\rm susy}^2\, M_{\rm Planck}\, (S + S^*) \; ,
\eeq
thus reintroducing a hierarchy problem. 
The values of $\Xi_F$ and $\Xi_S$ depend on the loop order at which the tadpoles are generated, which in turn depends on the particular non-renormalisable terms that give rise to the tadpoles. A solution to both the domain wall and the stability problem is to impose a discrete $R$-symmetry on the complete theory (including non-renormalisable operators) such that the tadpole terms are generated at high loop order~\cite{Panagiotakopoulos:1998yw}. One then obtains effective tadpole terms
\beq
\delta W = \xi_F\, S \; , \quad \delta V = \xi_S\, (S + S^*) \; , \quad {\rm where} \;\;\; \xi_F \lesssim M_{\rm susy}^2 \;\; {\rm and} \;\;\; \xi_S \lesssim M_{\rm susy}^3 \; .
\eeq
In the case where $\xi_F \sim M_{\rm susy}^2$ and $\xi_S \sim M_{\rm susy}^3$ the singlet cubic self interaction in the superpotential (\ref{eq:WNMSSM}) is not even phenomenologically required and can be omitted~\cite{Panagiotakopoulos:1999ah}. The resulting model has been denoted as the new MSSM or nMSSM as, in the limit where SUSY is unbroken, the MSSM $\mu$ term is only traded for the dimensionless $\l$ coupling. Once SUSY is softly broken, the generated tadpole terms $\xi_F$ and $\xi_S$ break both the $Z_3$ and the PQ symmetry. The superpotential of the nMSSM then reads
\beq\label{eq:WnMSSM}
W_{\rm nMSSM} = \l \widehat{S}\widehat{H}_u\widehat{H}_d + \xi_F\widehat{S}+ h_u\widehat{Q}\widehat{U}^c\widehat{H}_u + h_d\widehat{Q}\widehat{D}^c\widehat{H}_d + h_e\widehat{L}\widehat{E}^c\widehat{H}_d
\eeq
and the corresponding soft SUSY breaking potential is given by
\bea\label{eq:VnMSSM}
V_\mathrm{nMSSM} &=&
m_{H_u}^2 | H_u |^2 + m_{H_d}^2 | H_d |^2 + m_{S}^2 | S |^2 + \left( \l A_\l H_u H_d S + \xi_S S + \mathrm{h.c.}\right) \nn \\
&& +\, m_Q^2|Q^2| + m_U^2|U_R^2| + m_D^2|D_R^2| + m_L^2|L^2| + m_E^2|E_R^2| \nn \\
&& + \left( h_u A_u Q H_u U_R^c - h_d A_d Q H_d D_R^c - h_e A_e L H_d E_R^c + \mathrm{h.c.}\right) \nn \\
&& +\, M_1\, \widetilde{B}\, \widetilde{B} \, + \, M_2\, \widetilde{W}\, \widetilde{W} \, + \, M_3\, \widetilde{g}\, \widetilde{g} \; .
\eea

In this paper we study the general nMSSM with arbitrary soft terms at the SUSY scale as well as the semi-universal nMSSM for which one imposes the following constraints on the soft terms at the GUT scale
\beq
\left\{\ba{l}
m_Q = m_U = m_D = m_L = m_E \equiv m_0 \\
A_u = A_d = A_e \equiv A_0 \\
M_1 = M_2 = M_3 \equiv M_{1/2} \; .
\ea\right.
\eeq
In both cases one can trade the (free) parameters $m_{H_u}, m_{H_d}, m_{S}$ for the Higgs VEVs $v_u, v_d, s$, or equivalently for $\mu \equiv \l s$, $\tb \equiv \ds\frac{v_u}{v_d}$ and the known value of $M_Z^2 = g^2 v^2$, where $g^2 = (g_1^2 + g_2^2)/2$\,, $v = \sqrt{v_u^2 + v_d^2} \approx 174$~GeV, and $g_1$, $g_2$ denote the $U(1)_Y$ and $SU(2)_L$ gauge couplings, respectively.

From the SUSY $F$- and $D$-terms and the soft SUSY breaking terms one obtains the potential for the neutral Higgs fields:
\bea
V_\mathrm{Higgs} & = & \left| - \l H_u^0 H_d^0 + \xi_F \right|^2 + \frac{g^2}{4}\left(\left|H_u^0\right|^2 - \left|H_d^0\right|^2\right)^2 + m_{S}^2 |S|^2 \nn \\
&& +\, \left(m_{H_u}^2 + \left|\l S\right|^2\right) \left|H_u^0\right|^2 + \left(m_{H_d}^2 + \left|\l S\right|^2\right) \left|H_d^0\right|^2 \nn \\
&& + \left(- \l A_\l H_u^0 H_d^0 S + \xi_S S + \mathrm{h.c.}\right) \; ,
\eea
which at the minimum is
\bea
V_0 & = & \left(-\l v_u v_d + \xi_F\right)^2 + \frac{g^2}{4}\left(v_u^2 - v_d^2\right)^2 + m_{S}^2\, s^2 \nn \\
&& +\, \left(m_{H_u}^2 + \mu^2\right) v_u^2 + \left(m_{H_d}^2 + \mu^2\right) v_d^2 - 2 \l A_\l v_u v_d s + 2\xi_S s \; .
\eea
The minimisation equations are given by
\beq\label{eq:min}
\left\{\ba{l}
v_u \left(m_{H_u}^2 + \mu^2 + \l^2\,v_d^2 +\frac{g2}{2}(v_u^2-v_d^2)\right) - v_d \left(\mu A_\l +\l\xi_F\right)= 0 \; , \\
v_d \left(m_{H_d}^2 + \mu^2 + \l^2\,v_u^2 +\frac{g2}{2}(v_d^2-v_u^2)\right) - v_u \left(\mu A_\l +\l\xi_F\right)= 0 \; , \\
s\left(m_{S}^2 + \l^2(v_u^2+v_d^2)\right) + \xi_S - \l A_\l v_u v_d = 0 \; .
\ea\right.
\eeq
From the first two of these equations one can derive
\beq
\frac{v_u v_d}{v^2} \equiv \frac{1}{2}\sin 2\b = \frac{\mu A_\l +\l\xi_F} {m_{H_u}^2+m_{H_d}^2+2\mu^2+\l^2\,v^2} \; ,
\eeq
and from the third, one obtains in the limit where $s \gg v$ (or equivalently $\l \ll 1$)
\beq
\mu \simeq -\, \frac{\l\xi_S}{m_S^2} \; .
\eeq
In the basis $(H_{dR}, H_{uR}, S_R)$ and after the elimination of $m_{H_d}^2$, $m_{H_u}^2$ and $m_{S}^2$ using the minimisation equations (\ref{eq:min}), the elements of the $3 \times 3$ CP-even mass matrix ${\cal M}_S^2$ read 
\bea
{\cal M}_{S,11}^2 & = & g^2 v_d^2 + (\mu A_\l +\l\xi_F)\,\tb \; , \nn \\
{\cal M}_{S,22}^2 & = & g^2 v_u^2 + (\mu A_\l +\l\xi_F)/ \tb \; , \nn \\
{\cal M}_{S,33}^2 & = & \frac{\l^2 A_\l v_u v_d - \l\xi_S}{\mu} \; , \nn \\
{\cal M}_{S,12}^2 & = & (2\l^2 - g^2) v_u v_d - (\mu A_\l + \l\xi_F) \; , \nn\\ 
{\cal M}_{S,13}^2 & = & \l (2 \mu v_d - A_\l v_u) \; , \nn \\
{\cal M}_{S,13}^2 & = & \l (2 \mu v_u - A_\l v_d) \; .
\eea
Dropping the Goldstone mode, one can express the $2 \times 2$ CP-odd mass matrix ${\cal P}_S^2$ in the basis (${A}, S_I$), where ${A} = \cos\b\, H_{uI}+ \sin\b\, H_{dI}$
\beq
{\cal M}_{P,11}^2 = \frac{2 (\mu A_\l + \l\xi_F)}{\sin 2\b} \; , \quad
{\cal M}_{P,22}^2 = \frac{\l^2 A_\l v_u v_d - \l\xi_S}{\mu} \; , \quad
{\cal M}_{P,12}^2 = \l A_\l v \; .
\eeq
One can notice that ${\cal M}_{S,33}^2 = {\cal M}_{P,22}^2$, {\it i.e.} in the limit of small mixing between the singlet and doublet sectors, the CP-even and CP-odd singlet states have the same mass (up to radiative corrections, see below). In addition, this common (tree level) mass depends on the tadpole parameter $\xi_S$ which is a free parameter. Hence singlet like Higgs masses are arbitrary. In particular, they can be lighter than 125~GeV and still not excluded if their reduced couplings to SM particles (especially gauge bosons) are sufficiently suppressed.

Finally, in the basis $\psi^0 = (-i\l_1, -i\l_2^3, \psi_d^0, \psi_u^0, \psi_S)$, the neutralino mass matrix reads
\beq
{\cal M}_0 =
\left( \ba{ccccc}
M_1 & 0 & -\frac{g_1 v_d}{\sqrt{2}} & \frac{g_1 v_u}{\sqrt{2}} & 0 \\
& M_2 & \frac{g_2 v_d}{\sqrt{2}} & -\frac{g_2 v_u}{\sqrt{2}} & 0 \\
& & 0 & -\mu & -\l v_u \\
& & & 0 & -\l v_d \\
& & & & 0
\ea \right) \ .
\eeq
Therefore, in the limit of small singlino-higgsino mixing ($\l \ll 1 $), the singlino mass is $m_{\widetilde S} = 0$. On the other hand, if either $\mu$ or $M_1$ and $M_2$ are much larger than $M_Z$, one gets~\cite{Hesselbach:2007te}
\beq\label{eq:singlino}
m_{\widetilde S} \simeq \frac{\mu \l^2 v^2}{\mu^2 + \l^2 v^2}\, \sin 2\b\; .
\eeq
The experimental lower bound on $\mu$ (from the non observation of a light chargino) and the theoretical upper bound on $\l$ (assuming perturbativity up to the GUT scale) therefore yield an upper bound on the singlino mass $m_{\widetilde S} \lesssim 75$~GeV.

The physical CP-even Higgs states will be denoted as $h_i, i = 1,2,3$ (ordered in mass), and the physical CP-odd Higgs states as $a_i, i = 1, 2$. The neutralinos are denoted as $\tilde\chi^0_i, i = 1\dots5$ and their mixing angles $N_{i,j}$ such that $N_{1,5}$ indicates the singlino component of the lightest neutralino $\tilde\chi^0_1$ (assumed to be the LSP).

All the above expressions are for tree level mass matrices. Loop corrections play an important role, especially in the Higgs sector where they account for a large part of the SM like Higgs mass at 125~GeV. To compute the SUSY and Higgs spectrum, we have used the \nmssmtools\ package, setting the precision for radiative corrections to the minimum (precision flag for Higgs calculations = 0 in the \nmssmtools\ input files). This includes the full one loop and the leading log two loop contributions from (s)top/(s)bottom, as well as the leading log one loop EW corrections. We have not used the most precise computation (precision flag for Higgs calculations = 2) of ref.~\cite{Degrassi:2009yq}, as it is valid only for the $Z_3$ invariant NMSSM. In this ($Z_3$ invariant) limit however, we have checked that the difference between the two computations is usually $\lesssim 3$~GeV for the SM like Higgs state near 125~GeV. In addition, a slight change of input parameters can always reproduce the same Higgs spectrum with both flags~\footnote{For a review of NMSSM Higgs mass calculations in public codes (including \nmssmtools), see ref.~\cite{Staub:2015aea}.}. The minimal precision for radiative corrections presents the extra advantage of using less CPU time, which is crucial for scans on large parameter space. In addition it allows to compute easily the complete effective Lagrangian in the Higgs sector with the same level of approximation. This complete Lagrangian can then be fed into \micro\ so as to compute the relic density of the LSP DM candidate as well as its DD and ID rates. Note that the higher-order corrections to the Higgs self-couplings encoded in the effective Lagrangian can in some cases have a significant effect on the DM relic density.

\section{Parameter scan}
\label{sec:sec3}

The parameter exploration of the semi-universal nMSSM (as defined in Sec.~\ref{sec:sec2}) has been carried out using \nmssmtools~\verb#v4.6.0#, scanning over the following parameters~\footnote{The value of the top quark pole mass has been fixed to $m_{\rm top} = 173.1$~GeV. }:
\beq
m_0, \; M_{1/2}, \; A_0, \; \mu, \; \tb, \; \l, \; \xi_F, \; \xi_S, \; A_\l,
\eeq
which are all defined at the GUT scale except $\tb$ (at $M_Z$) and $\l$, $\mu$ (at the SUSY scale). To efficiently scan over the nMSSM parameter space we have employed the Markov Chain Monte Carlo (MCMC) routines implemented in the \nmssmtools\ package, which we have tuned in order to cover in details regions of parameter space corresponding to lighter sparticles, {\it i.e.} with higher experimental prospects. Scenarios with very heavy sparticles (out of the LHC Run-2 reach) have been discarded.

We have applied all the default constraints implemented into \nmssmtools\ (except for the constraint on the anomalous magnetic moment of the muon g$_{\mu}-2$), which include in particular~\footnote{See \url{http://www.th.u-psud.fr/NMHDECAY/nmssmtools.html} for a detailed list of the implemented constraints.}:
\begin{itemize}
 \item[-] No unphysical minimum of the Higgs potential,
 \item[-] No Landau pole below the GUT scale,
 \item[-] Invisible Z width $\Gamma_Z < 0.5$~MeV,
 \item[-] $B$-physics constraints,
 \item[-] LEP and Tevatron searches for sparticles and Higgs bosons,
 \item[-] Tevatron and LHC searches on charged Higgs via top decays,
 \item[-] At least one Higgs boson in the $125.1\pm3$~GeV mass range,
 \item[-] $\chi^2$ fit to the Higgs signal strengths~\cite{Belanger:2013xza}.
\end{itemize}
The latter indirectly takes into account the limit on non-standard decays of the SM like Higgs, such as the decay into light Higgs states or the invisible decays into the LSP which are somewhat dependent on shifts of other Higgs couplings. Moreover we have checked a posteriori that the direct limits on heavy Higgs states in the $WW$ channel were satisfied~\cite{Aad:2015agg}.

In performing our scan we have also required the DM relic density $\Omega h^2$ to be compatible with the relic abundance measured by Planck~\cite{Adam:2015rua}, $\Omega h^2_{\rm Planck}= 0.1186\pm0.0020$ at 68\% CL.
We have chosen to impose just an upper bound of the relic density, $\Omega h^2<0.131$, which takes into account $\sim$ 10\% theoretical uncertainties that could arise from loop corrections into the DM annihilation cross section, see {\it e.g.}~\cite{Baro:2007em}.
We have also required the spin independent cross section for DD, rescaled for the local DM abundance ($\sigma^{\rm SI}_{\rm rescaled}=\sigma^{\rm SI} \Omega h^2/\Omega h^2_{\rm Planck}$), to be compatible with the latest LUX results~\cite{Akerib:2013tjd}.

\begin{figure}[htb]
\centering
\includegraphics[width=0.46\textwidth]{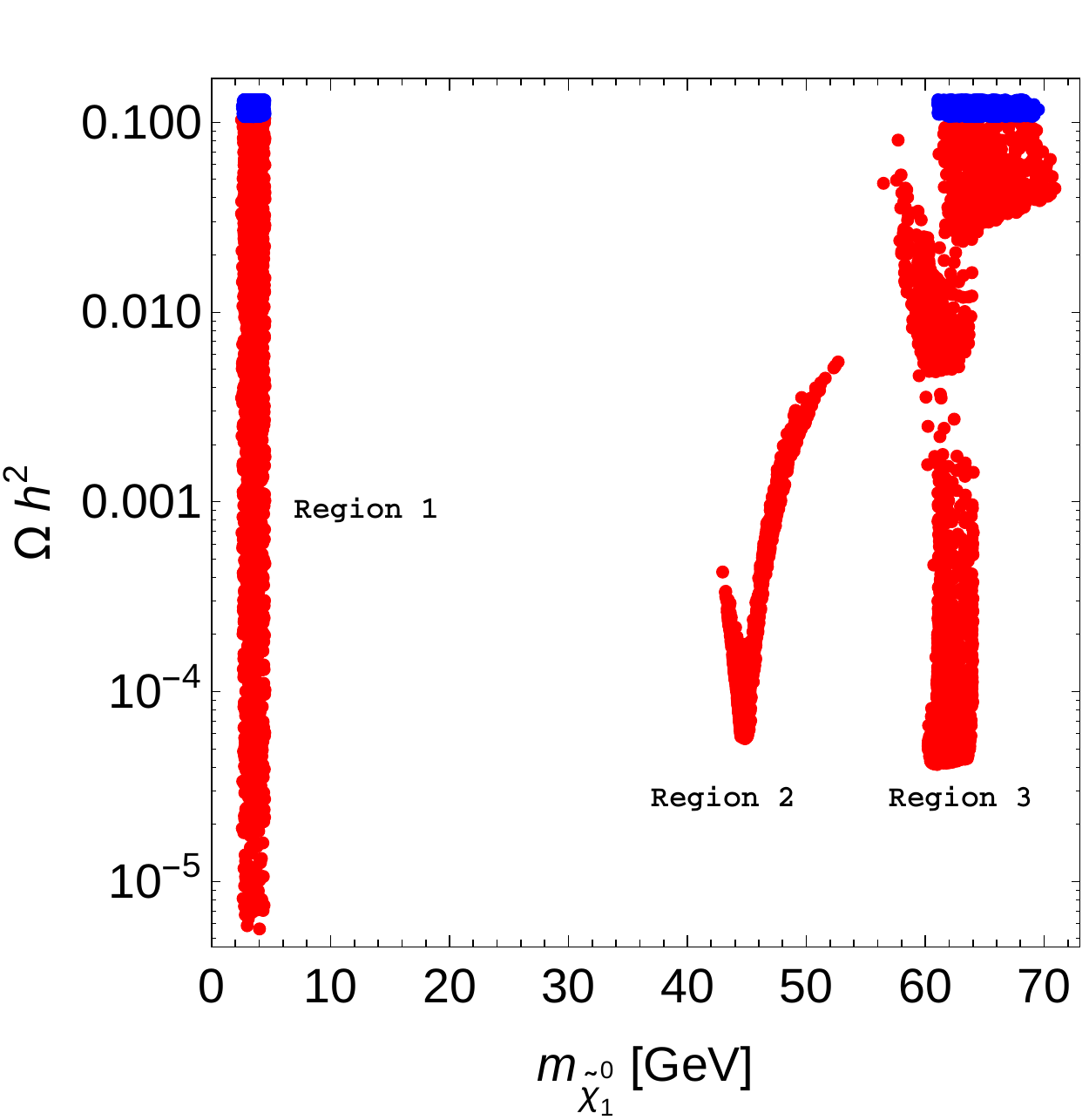}
\caption{DM relic density $\Omega h^2$ in function of the LSP mass. Blue (red) points correspond to a DM relic density $0.107\le\Omega h^2\le 0.131$ ($\Omega h^2<0.107$).}
\label{fig:scan_res}
\end{figure}

We illustrate in Fig.~\ref{fig:scan_res} the results of the scan mapped in the $m_{\tilde\chi^0_1}$ - $\Omega h^2$ plane, showing in blue the points with a DM relic density compatible with Planck (0.107$<\Omega h^2<$0.131) and in red the points for which it is below ($\Omega h^2<$0.107). We see that three different regimes for the LSP mass exist: a region with a very light LSP below 5~GeV (region 1), a region with a $\sim$~45~GeV LSP (region 2) and a region with a $\sim 65$~GeV LSP (region~3). In the first region the DM annihilation proceeds through a light pseudoscalar (singlet like) Higgs resonance, the second corresponds to the $Z$ resonance and the third to the exchange of a SM like Higgs or $Z$ boson. Note that the gap for neutralino masses between 5 and 40~GeV is mostly due to constraints from the invisible width of the Higgs. The light LSP is a nearly pure singlino, so that the Higgs invisible width is very small. As the singlino mass increases, the same does its higgsino component, hence increasing the contribution to the Higgs invisible width. 

As we will discuss in the following, the 5~GeV and 65~GeV LSP region present two and three sub-regions respectively, mapped in different areas of the $m_0$-$ M_{1/2}$ parameter space. We report in Tabs.~\ref{tab:scan-res-par1}-\ref{tab:scan-res-par1_bis} the maximum and minimum values for all of the nMSSM input parameters, in the three regions and sub-regions that we have identified. Moreover in this Tables we also specify the weak scale gaugino masses and give a quick overview of the sparticle spectrum in the different sub-regions. All regions have small $\mu$ and several sub-regions feature sfermions and/or gauginos at the multi TeV scale. We distinguish region 1 for which the 125~GeV (SM like) Higgs state is the second lightest CP even ($h_2$) while the lightest ($h_1$) is mainly singlet, from regions 2 and 3 where the lightest CP even Higgs state ($h_1$) is SM like with a mass of 125~GeV. In the former case (with intermediate values of $\tb$) the SM like Higgs state takes its (relatively heavy) mass from $h_1 / h_2$ mixing effects. In the latter, $\tb \approx 2$ and $\l$ is near its maximal theoretically allowed value (from perturbativity up to the GUT scale). Hence the pure nMSSM contribution to the lightest Higgs mass is maximal.

\begin{table}
\begin{center}
\begin{tabular}{| l || c | c || c |}
\hline
Region & 1A & 1B & 2 \\
\hline
\hline
$\tb$ & 6.6 ~~~ 10 & 6 ~~~ 8 &1.5 ~~~ 2.1 \\
$\l$ & 0.33 ~~~ 0.53 & 0.49 ~~~ 0.52 & 0.68 ~~~ 0.80 \\
$\mu$ & 240 ~~~ 400 & 350 ~~~ 430 & 115 ~~~ 180 \\
$m_0$ & 0 ~~~ 1080 & 4040 ~~~ 4800 & 2$\cdot 10^4$ ~~~ 6$\cdot10^5$ \\
$ M_{1/2}$ & 630 ~~~ 1200 & 280 ~~~ 440 & 180 ~~~ 470 \\
$A_0$ & $-$1700 ~~~ 50 & 6700 ~~~ 7900 & $-$3.7$\cdot10^4$ ~~~ $-$2.5$\cdot10^3$ \\
$A_\l$ & 1400 ~~~ 6000 & 7000 ~~~ 7900 & $-$1.3$\cdot10^4$ ~~~ 3.3$\cdot10^4$ \\
$\xi_F$ & 10 ~~~ 100 & $-$1.5$\cdot10^4$ ~~~ $-$1.4$\cdot10^4$ & 3.7$\cdot10^4$ ~~~ $5.1\cdot10^6$ \\
$\xi_S$ & $-$6$\cdot10^4$ ~~~ 2$\cdot10^4$ & $-$1.9$\cdot10^7$ ~~~ $-$1.6$\cdot10^7$ & $-$5.2$\cdot10^{10}$ ~~~ 9.7$\cdot10^8$ \\
\hline
\hline
$M_1$ & 270 ~~~ 520 & 110 ~~~ 190 & 95 ~~~ 225 \\
$M_2$ & 500 ~~~ 950 & 200 ~~~ 340 & 160 ~~~ 400 \\
\hline
\hline
$m_{\tilde q}$ & 1300 ~~ 2400 & $>$ 3000 & $>$ 20000 \\
$m_{\tilde t_1}$ & 350 ~~ 1300 & 1050 ~~ 1900 & $>$ 3000 \\
$m_{\tilde l}$ & 180 ~~ 1100 & $>$ 3000 & $>$20000 \\
$m_{\tilde g}$ & 1450 ~~ 2600 & 780 ~~ 1250 & 800 ~~ 1500 \\
\hline 
\end{tabular}
\caption{Maximum and minimum parameter values surviving the constraints implemented in \nmssmtools\ for regions 1 and 2. Also indicated are the $M_1$ and $M_2$ values at the SUSY scale and the mass ranges for sparticles. Dimensionfull parameters are expressed in GeV except $\xi_F$ and $\xi_S$ which are in GeV$^2$ and GeV$^3$ respectively.}
\label{tab:scan-res-par1}
\end{center}
\end{table}

\begin{table}
\begin{center}
\vskip 6pt
\begin{tabular}{| l || c | c | c |}
\hline
Region & 3A & 3B & 3C \\
\hline 
\hline 
$\tb$ & 1.8 ~~~ 2.5 & 1.8 ~~~ 2.8 &1.3 ~~~ 1.8 \\
$\l$ & 0.64 ~~~ 0.77 & 0.66 ~~~ 0.74 & 0.65 ~~~ 0.74 \\
$\mu$ & $-$140 ~~~ $-90$ & $-$110 ~~~ $-$90 & 110 ~~~ 150 \\
$m_0$ & 3.9$\cdot 10^3$ ~~~ 6.6$\cdot10^4$ & 170 ~~~ 2500 & 0 ~~~ 3150 \\
$ M_{1/2}$ & 130 ~~~ 210 & 200 ~~~ 560 & 5.6$\cdot10^3$ ~~~ 2.3$\cdot10^4$ \\
$A_0$ & 11 ~~~ 3.3$\cdot10^4$ & 440 ~~~ 3600 & $-$1.9$\cdot10^4$ ~~~ 4.7$\cdot10^3$ \\
$A_\l$ & 6.2$\cdot 10^3$ ~~~ 2.7$\cdot10^4$ & 450~~~3500 & 4 ~~~ 8200 \\
$\xi_F$ & 4.8$\cdot10^5$ ~~~ 3.4$\cdot10^7$ & 4.3$\cdot10^5$ ~~~ 7.5$\cdot10^6$ & 4.6$\cdot10^5$ ~~~ $1.8\cdot10^7$ \\
$\xi_S$ & 2.0$\cdot10^9$ ~~~ 1.3$\cdot10^{11}$ & 1.0$\cdot10^8$ ~~~ 4.9$\cdot10^9$ & $-$1.9$\cdot10^{9}$ ~~~ 3.0$\cdot10^9$ \\
\hline
\hline
$M_1$ & 52 ~~~ 65 & 85 ~~~ 230 & 2.6$\cdot10^3$ ~~~ 1.1$\cdot10^4$ \\
$M_2$ & 83 ~~~ 108 & 160 ~~~ 430 & 4.6$\cdot10^3$ ~~~ 2.0$\cdot10^4$ \\
\hline 
\hline 
$m_{\tilde q}$ & $>$ 3000& 780~~2500& $>$ 10000\\
$m_{\tilde t_1}$ & 500~~$>$20000& 150~~550& $>$ 5000\\
$m_{\tilde l}$ & $>$3000& 100~~2500& 100~~3000\\ 
$m_{\tilde g}$ & 450~~650& 590~~1300& $>$ 10000\\ 
\hline 
\end{tabular}
\caption{Maximum and minimum parameter values surviving the constraints implemented in \nmssmtools\ for region 3. See caption of Tab.~\ref{tab:scan-res-par1}.}
\label{tab:scan-res-par1_bis}
\end{center}
\end{table}

In the next Section we will analyse these three regions separately, first checking what are the constraints set by LHC Run-1 and ID experiments, then describing the prospects of this scenarios for the LHC Run-2, together with the prospects for future DD experiments.

To check the compatibility with the LHC constraints we have used both the packages \smodels~\verb#v1.0.2# and \ma5.
The former is a tool designed to decompose the signal of a given SUSY model into simplified topologies that are searched for by ATLAS and CMS, taking into account that in a generic SUSY spectrum the assumptions on the sparticles decays can be (and usually are) different from the ones assumed in the experimental searches. The input of \smodels\ consists of an \verb#SLHA# file~\cite{Allanach:2008qq}, containing particle spectrum and decay tables, together with SUSY production cross sections, which we have computed with \micro~\verb#v4.1.7#. For strong production of sparticles we have also applied next-to-leading-order and next-to-next-to-leading-log (NLO+NNLL) QCD $\kappa$-factors, which have been computed with the \verb#nllfast# package~\cite{Beenakker:1996ch,Beenakker:1997ut,Kulesza:2008jb,Kulesza:2009kq,Beenakker:2009ha,Beenakker:2010nq,Beenakker:2011fu}. Note that due to the presence of many states below the TeV scale, and in particular of a very light LSP, for some regions of parameter space the coloured sparticles possess a large number of possible decay chains, including long ones. Hence building the list of existing topologies can be much more computer-time consuming than in the MSSM. Conversely \ma5 is a multi purpose package designed for the analysis of events generated at parton level and/or reconstructed level with the inclusion of parton showering and detector effects. In this work we have exploited the \ma5 Public Analysis Database, {\it i.e.} a list of implemented and validated experimental searches, with which one can recast the experimental limits set by ATLAS and CMS in a generic model or in a particular model configuration. Clearly the approach of \ma5 is more general, since it does not rely on any simplified model assumption, but it has the drawback of being somewhat more time consuming, given the need to generate events with a Monte Carlo generator for each model point that one wants to test. In this paper we will use a combination of both tools to check the limits set by LHC Run-1 in the nMSSM, using \ma5 especially to set bounds on the gluino mass.

\section{LHC and DM phenomenology}
\label{sec:sec4}

After having defined the different regions of the GUT scale parameter space obeying the basic set of collider, astrophysical, cosmological and theoretical constraints, we now examine for each scenario the constraints from searches for sparticles at LHC Run-1 as well as from indirect searches for DM. For each scenario we then discuss what are the most relevant searches at LHC Run-2, both for sparticles and Higgs states, and we provide benchmark points with specific nMSSM signatures. The complementarity with direct DM searches is also highlighted.

\subsection{Region 1, $m_{\tilde\chi^0_1}<$ 5~GeV}

In this region the LSP is a quasi pure singlino state and it can appear in two different configurations of the $m_0$-$ M_{1/2}$ parameter space. One is characterised by small $m_0$~($\ltap$~1~TeV) and $ M_{1/2}\sim$~1~TeV (region 1A) while the second one has small $ M_{1/2}$ ($\ltap$~500~GeV) and large $m_0$~$\sim$~4~TeV (region 1B). These two regions give rise to different sparticle spectra that we will analyse separately. We however stress three common features in these sub-regions. The first is that the lightest pseudoscalar and the singlino masses are linked, with $|m_{a_1}-2m_{\tilde\chi^0_1}|\ltap1.5$~GeV, in order to ensure efficient annihilation of the singlino through $a_1$ resonance, so as to satisfy the relic density constraint. The second follows from the singlino nature of the LSP. Decays of sparticles to a final state containing the LSP will often be suppressed thus, phase space allowing, longer decay chains involving heavier neutralinos, such as $\tilde t_1\to t \tilde\chi_2^0\to t \tilde\chi^0_1 Z$, will be preferred. This has important consequences for simplified models limits. Finally, the measured DM relic abundance can be accounted for just by the LSP itself. However, as explained in the previous Section, we will work in the less constraining assumption of a relic abundance $\Omega h^2<$ 0.131, thus allowing for the possibility of another DM component.

\subsubsection{Current constraints}
\label{sec:1AB_constraints}

In the region with small $m_0$ (1A) the spectrum is characterised by the presence of light stops ($m_{\tilde t_1}< 1300$~GeV) and light sleptons ($m_{\tilde l}\sim 180$--$1100$~GeV), with first and second generation squarks heavier than 1.3~TeV and gluinos heavier than 1.45~TeV. The most stringent constraints on coloured sparticles are thus easily evaded. Given the values of $\mu, M_1$ and $M_2$ reported in Tab.~\ref{tab:scan-res-par1}, the EWino spectrum is characterised, besides the light singlino, by four neutral and charged states with a mass between 240 and 520~GeV corresponding to the three higgsinos and the bino. In the Higgs sector the role of the SM Higgs boson is played by $h_2$, while the $h_1$ mass is between 35 and 70~GeV and the lightest pseudoscalar $a_1$ has a mass $\sim$ 2 $m_{\tilde\chi^0_1}$. The masses of the three states of the heavy Higgs doublet are above 1.5~TeV. Note that pure scalar and pseudoscalar singlets are expected to have the same mass, however the mixing with the light scalar doublet is sufficient to increase the mass of the scalar singlet by a few tens of GeV.

\begin{figure}[htb]
\begin{center}
\includegraphics[width=0.46\textwidth]{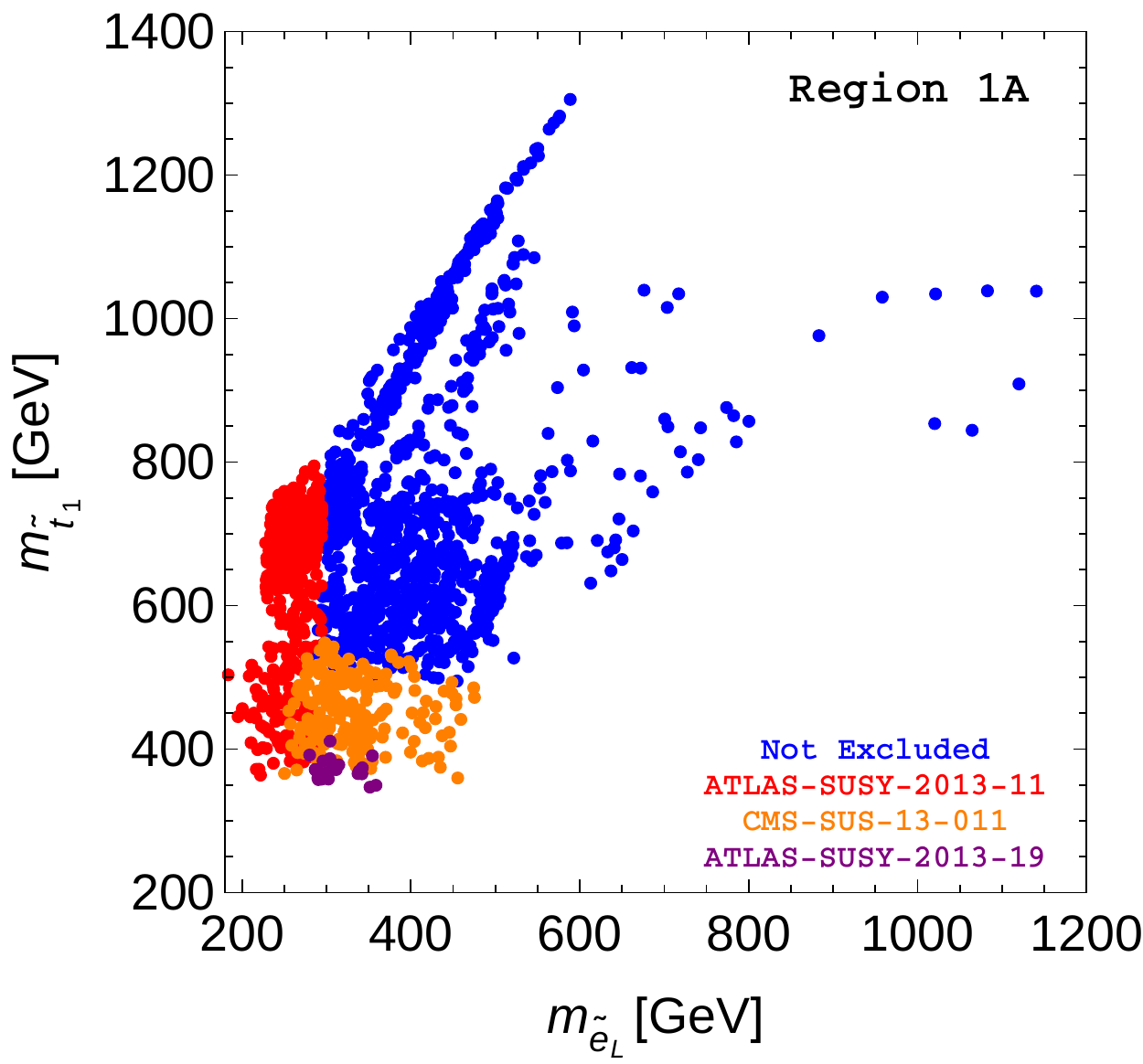}
\caption{Allowed and excluded points for region 1A in the $m_{\tilde e_L}$-$m_{\tilde t_1}$ plane. For each excluded point we indicate the search with the maximum sensitivity.}
\label{fig:1A_smodels}
\end{center}
\end{figure}

We show in Fig.~\ref{fig:1A_smodels} the allowed and excluded points after the application of the constraints implemented in \smodels, projected in the $m_{\tilde e_L}$-$m_{\tilde t_1}$ plane. For each excluded point we indicate the most constraining analysis. This is defined in \smodels~as the analysis that has the larger ratio between the theoretical prediction and the experimental measurement for a given channel. However more than one channel can exclude the same point. From Fig.~\ref{fig:1A_smodels} it is clear that the most constraining searches are the ATLAS search for slepton production~\cite{Aad:2014vma} ($\tilde l\to l \tilde\chi^0_1$) and the CMS search for stop production~\cite{Chatrchyan:2013xna} ($\tilde t\to b \tilde\chi^+_1$ and $\tilde t\to t \tilde\chi^0_1$). In particular the former is able to exclude sleptons lighter than $\sim 300$~GeV while the latter sets a lower bound of $\sim 550$~GeV on the lightest stop mass. These are stringent bounds, in the sense that no sparticles lighter than these limits are allowed. It is important to notice that, while the reach on the slepton mass is close to the official result of the ATLAS analysis for a 5~GeV LSP ($m_{\tilde l}\gtap 330$~GeV), this is not the case for the stop search, where the ATLAS limit is around 650~GeV. However this can be explained by the fact that the simplified model result assumes a 100\% branching ratio either for $\tilde t_1\to t \tilde\chi^0_1$ or $\tilde t_1\to b \tilde\chi^+_1$, an assumption which is not satisfied here. First, the mass spectrum is such that there is always at least one decay channel into a heavier neutralino which is allowed. Moreover, the decays into heavier neutralinos typically have larger branching ratios than the decay into the singlino LSP. This causes therefore a small reduction of the LHC exclusion reach.

The second region with a light LSP (1B) corresponds to large $m_0$ ($\sim 4$~TeV) and small $ M_{1/2}$. The large value of $m_0$ yields heavy sfermions with slepton and squark masses above 3~TeV, except for the lightest stop which is between 1--2~TeV. Therefore the only sub-TeV sparticles are EWinos and gluinos. In particular the neutralino spectrum has the following hierarchy: a light singlino ($m_{\tilde\chi^0_1}<$ 5~GeV), a bino ($m_{\tilde \chi_2^0}\sim 100$--$200$~GeV), a wino ($m_{\tilde\chi^0_3}\sim$~200--300~GeV) which is degenerate with the lightest chargino, and heavier higgsino states since typically $\mu$ is larger than $M_1, M_2$ in region 1A (see Tab.~\ref{tab:scan-res-par1}). The gluinos lie in the 800--1200~GeV range, while $h_1$ is heavier than in region 1A, between 70 and 90~GeV.

Contrary to what we found in region 1A, \smodels~does not set any constraint on the parameter space of this region. Clearly sleptons and squarks, including stops, escape the limits set by the experimental searches due to their high masses. Conversely, the reasons for EWinos escaping the LHC limits are less straightforward. The chargino mass in region 1B is typically above the bound set by ATLAS from the search for $\tilde\chi_1^+\tilde\chi^-_1$ production, with a subsequent $\tilde\chi^\pm_1\to W^\pm \tilde\chi^0_1$ decay, which, for a light LSP, is $m_{\tilde\chi^+} > 180$~GeV~\cite{Aad:2014vma}. The other simplified topology analysed by ATLAS which is relevant for region 1B is $\tilde\chi^\pm_1\tilde\chi^0_2\to W Z \tilde\chi^0_1 \tilde\chi^0_1$. It assumes 100\% EWinos decays into the LSP plus a SM gauge boson, as well as pure wino states ($m_{\tilde\chi^+_1} = m_{\tilde\chi^0_2}$). Considering the neutralino compositions, the latter criteria applies in our case only to $\tilde\chi^0_3$ while the decay assumption applies only to $\tilde\chi^\pm_1$. Indeed, the branching ratio for $\tilde\chi^0_3 \to Z\tilde\chi^0_1$ can be suppressed when decay rates into $h_{1,2}\tilde\chi^0_1$ and $a_1 \tilde\chi^0_1$ become important. The same comment can be made about the $\tilde\chi^0_2$ decays. Moreover the $\tilde\chi^0_2$ composition, which is mainly bino, causes a significant decrease of the $\tilde\chi^\pm_1\chi^0_2$ cross sections with respect to the wino case assumed in simplified models. Finally, the lack of sensitivity of the ATLAS search in a $3l+E_T^{\rm miss}$ final state~\cite{Aad:2014nua} can be explained by the small branching ratios of the neutralinos in lepton pairs and LSP, which is typical of scenarios with heavy sleptons. 

The case of the gluino deserves special consideration, since a priori its mass can be well within the range excluded by simplified models analyses. As before, the issue is that we find reduced branching ratios in the standard search channels ($\tilde g\to t\bar t\tilde\chi^0_1,t\bar b\tilde\chi^0_1$, $q\bar q\tilde\chi^0_1$) because of competing decays into heavier neutral and charged EWinos, that can once again be linked to the singlino nature of the LSP. However, gluino searches generally rely on a large number of jets in the final state. These signal regions can be easily fed also from gluino decays different from the standard assumptions, in particular decays into heavier neutralinos.

We have then used the recast~\cite{MA5:CMS-SUS-13-016} CMS gluino search~\cite{CMS:2013ija}, implemented in the package~\ma5, which relies on an opposite sign dilepton final state, more than 4 jets and more than 2 b-tag jets, together with a large $E_T^{\rm miss}$. This search sets a 95\%~CL exclusion when more than 4 events fall in its unique signal region.

We have simulated gluino pair production with \verb#MadGraph5 v1.5.11#~\cite{Alwall:2014hca}. Parton showering, hadronization and decay of unstable particles have been performed with the package \verb#Pythia v6.4#~\cite{Sjostrand:2006za}, while the \ma5 tuned version of \verb#Delphes v3.2.0#~\cite{deFavereau:2013fsa} has been used to simulate detector effects. Jets are reconstructed with \verb#FastJet#~\cite{Cacciari:2011ma}, via an anti-$k_T$~\cite{Cacciari:2008gp} algorithm.
We have then computed, through the efficiency map function implemented in \ma5, acceptances times efficiencies for various values of the gluino mass. From these, we have then calculated the final number of events for our nMSSM scenarios, using the production cross sections for gluino pair production computed via \micro\ and \verb#nllfast#. We have found a lower limit on the gluino mass of $\sim 1.1$~TeV, close to the official CMS result~\cite{CMS:2013ija}. This result strongly constrains region 1B, leaving just a few points with a heavy enough gluino.

\begin{figure}[htb]
\includegraphics[width=0.46\textwidth]{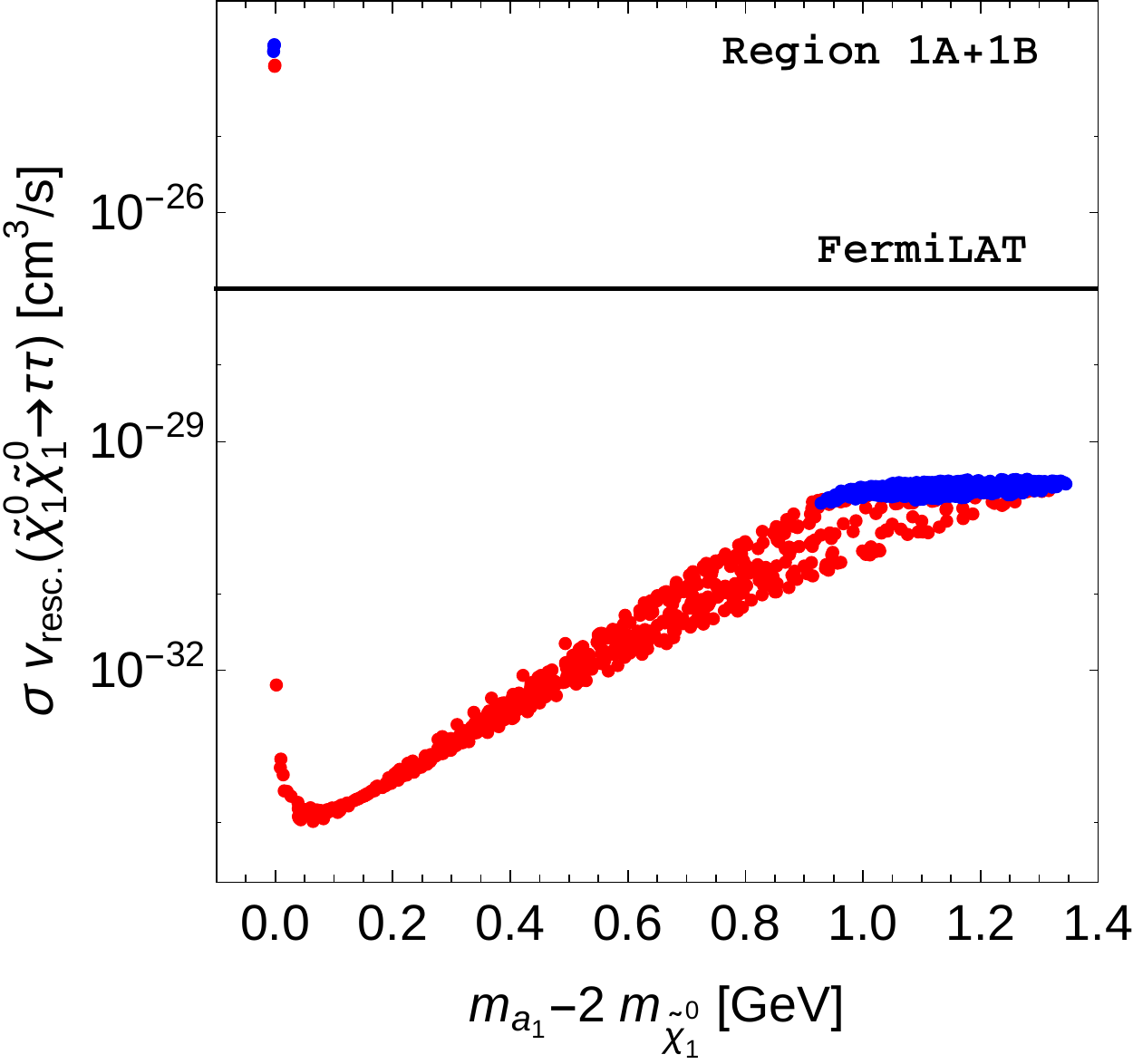}\hfill
\includegraphics[width=0.46\textwidth]{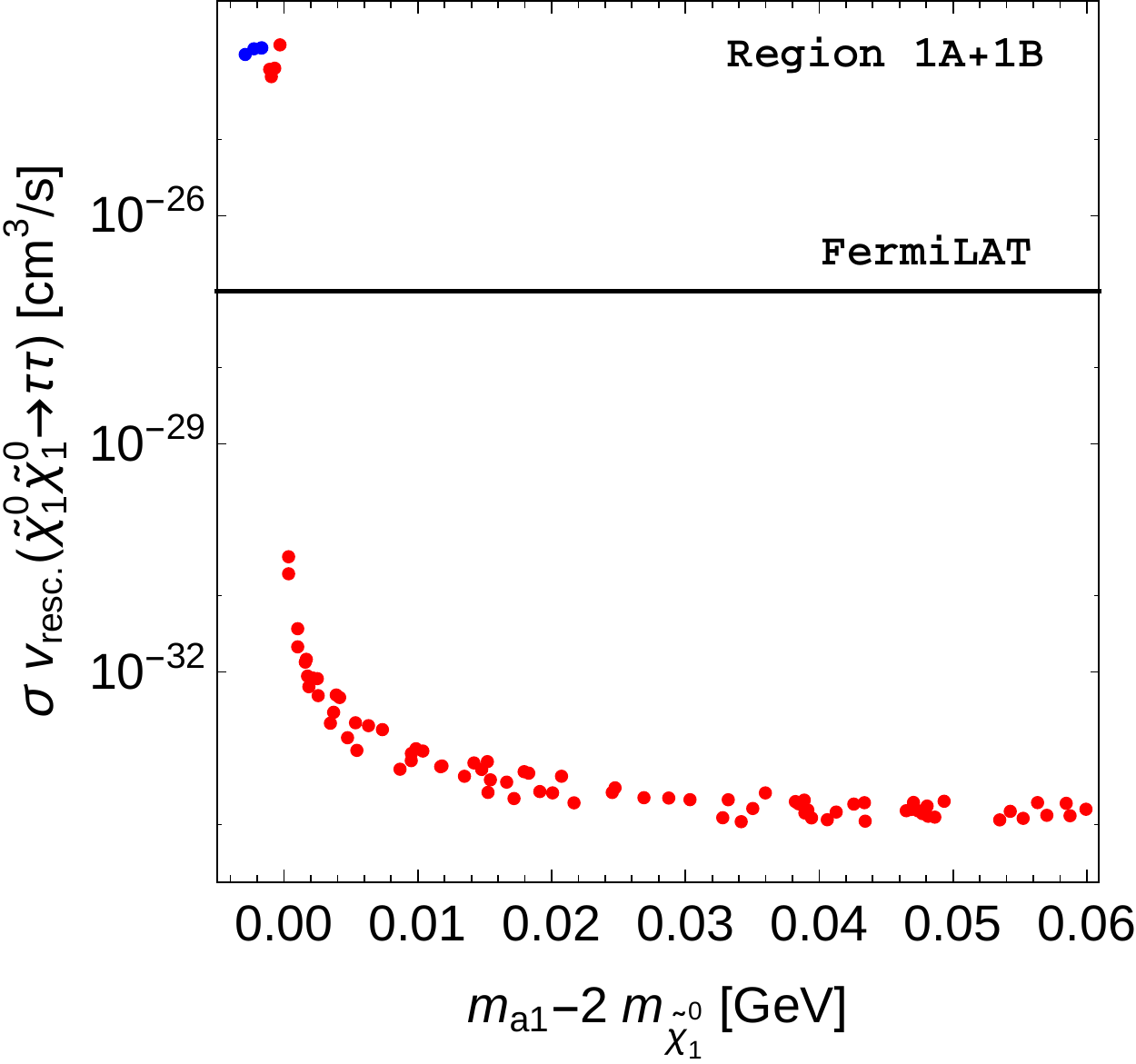}
\caption{ID detection rates into $\tau^+\tau^-$ final state in function of $m_{a_1}-2 m_{\tilde\chi^0_1}$. Blue (red) points correspond to a DM relic density $0.107\le\Omega h^2\le 0.131$ ($\Omega h^2<0.107$). The horizontal lines represent the current limit from Fermi-LAT. Regions 1A and 1B are shown together.}
\label{fig:1A_ID}
\end{figure}

Finally, we have also computed the ID cross section for LSP pair annihilation into $\tau^+\tau^-$ final state (the most relevant at these masses) rescaled by the square of the local DM density, $[\Omega h^2/ (\Omega h^2)_{\rm Planck}]^2$, and compared it with the exclusion limits set by Fermi-LAT~\cite{Ackermann:2015zua}. The results are shown in Fig.~\ref{fig:1A_ID}. For a $\sim 5$~GeV LSP the Fermi-LAT limit is approximately constant and equal to $10^{-27}{\rm cm^3/s}$. In the left panel we show all the points surviving the LHC Run-1 constraints (implemented as discussed above), while in the right panel we zoom on to the region with small $m_{a_1}-2 m_{\tilde\chi^0_1}$. Blue (red) points correspond to a DM relic density $0.107\le\Omega h^2\le 0.131$ ($\Omega h^2<0.107$). From the plots it is clear that the current results from Fermi-LAT are able to completely exclude the portion of parameter space which corresponds to $m_{a_1}<2 m_{\tilde\chi^0_1}$, while above this threshold ID detection rates are well below the experimental limits. The reason for this is simply that when $2m_{\tilde\chi^0_1}$ is just below $m_{a_1}$, its annihilation cross section at small velocities can be significantly enhanced by the resonance effect, while in the early universe the thermal velocity of the neutralino is enough to overshoot the resonance~\cite{AlbornozVasquez:2011js,Bi:2009uj}. 
The increase in the rescaled cross section with  $m_{a_1}-2 m_{\tilde\chi_1^0}>0.1~\rm{GeV}$ in Fig.~\ref{fig:1A_ID} (left panel) is an effect of the rescaling.
Indeed  $\sigma  v$ in the galaxy as well as in the early universe  decreases when  moving away from the narrow pseudoscalar resonance until one reaches a region where
$m_{a_1}-2 m_{\tilde\chi_1^0}$ is too  large to benefit from a strong resonance enhancement and the relic density is in agreement with PLANCK data. 
However since the rescaling factor is  inversely  proportional to the square of the annihilation cross section in the early universe, the net effect is an increase of the rescaled cross section for DD until $m_{a_1}-2 m_{\tilde\chi_1^0} \approx 1 $ GeV.
Note that limits from AMS antiproton are not expected to be important for such low masses~\cite{Cirelli:2013hv}.

\subsubsection{LHC Run-2 and direct detection experiments prospects}

We now discuss the prospects for the LHC Run-2, including sparticles and Higgs searches, as well as DD experiments for the points surviving the constraints imposed by the LHC Run-1 and ID measurements. 

As shown in Fig.~\ref{fig:1A_smodels} standard searches for stops quarks are already quite effective in constraining light stops in region 1A, for which the dominant decay mode is $\tilde t_1\to b\tilde\chi^+_1 $, and Run-2 is therefore expected to significantly extend the reach in this channel. The standard search channel $\tilde t_1\to t\tilde\chi^0_1$ does not offer as good prospects, since the branching ratio into the singlino LSP is typically suppressed. For example we found a 600~GeV stop with a $\sim$~60\% branching ratio into a chargino and only $\sim$ 5\% in $t \tilde\chi^0_1$, with the rest of the decays being saturated by $t\tilde\chi^0_{2,3}$. Since $\tilde\chi^0_{2,3}$ mainly decay into a $Z$ or a SM like Higgs boson and a LSP, final states with $tZ+E_T^{\rm miss}$ or $th_2+E_T^{\rm miss}$, arising from one or both of the pair produced stops, provide a characteristic signature in this region. Also slepton searches will clearly act as a powerful probe of this region. In fact, only a factor two improvement in the LHC Run-2 mass reach, relative to the one of Run-1, would cover most of region 1A, leaving only a few points with sleptons heavier than 500~GeV. Upgrades of standard EWinos searches, as in ref.~\cite{Aad:2014nua}, will also be able to further extend the LHC reach. In fact, even if in Fig.~\ref{fig:1A_smodels} the strongest bound represented is the one arising from slepton pair production, the decay pattern of $\tilde\chi^+_1$ and $\tilde\chi^0_2$ can be consistent with the assumption in the experimental searches ({\it e.g.} $\tilde\chi^0_2\to \tilde\chi^0_1 Z$ and $\tilde\chi^\pm_1\to \tilde\chi^0_1 W$), the latter especially if sleptons are heavier than EWinos. Conversely, for region 1B, gluino searches will clearly represent the major probe for LHC Run-2, since $m_{\tilde g}<1.2$~TeV and exclusion limits on the gluino mass are expected to greatly increase already with early 13~TeV data.

The Higgs sector of these regions is characterised by a light pseudoscalar $a_1$, accompanied by a light scalar $h_1$, both with a dominant singlet component. The mass of $h_1$ lies between 30--70~GeV in region 1A and 70--90~GeV in region 1B. This leads, in region 1A, to the interesting possibility of exotic decays of the 125~GeV (SM like) Higgs state $h_2$ into a pair of light scalars ($h_2\to h_1 h_1$) as well as pseudoscalars ($h_2\to a_1 a_1$). Given the mass of the $a_1$ and of the $h_1$, possible interesting signatures to be explored are $h_2\to a_1 a_1\to 4 \tau/ 2\tau 2 \mu$ and $h_2\to h_1 h_1\to 4b/2b 2\tau$. 

\begin{figure}[htb]
\includegraphics[width=0.46\textwidth]{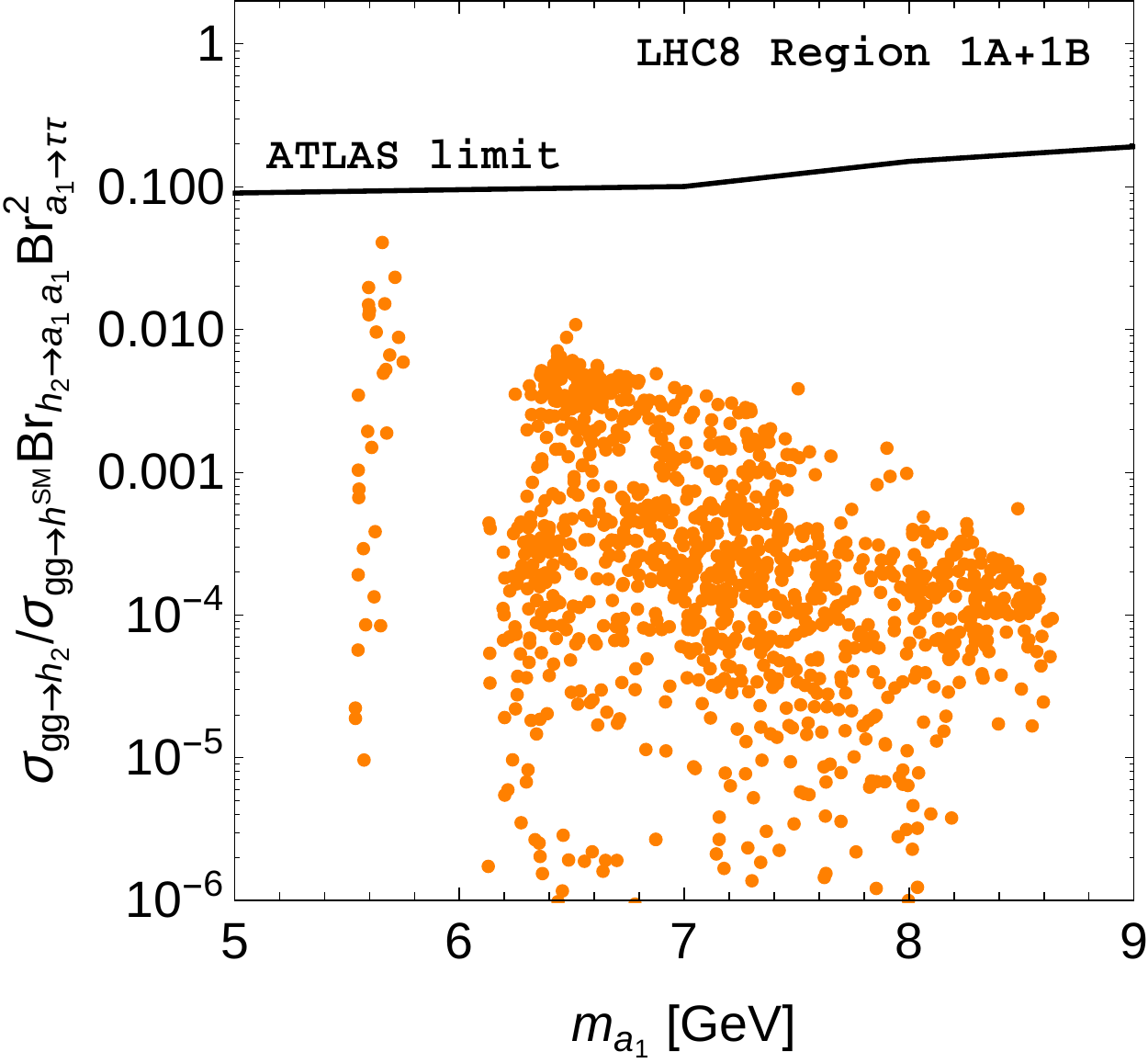}\hfill
\includegraphics[width=0.46\textwidth]{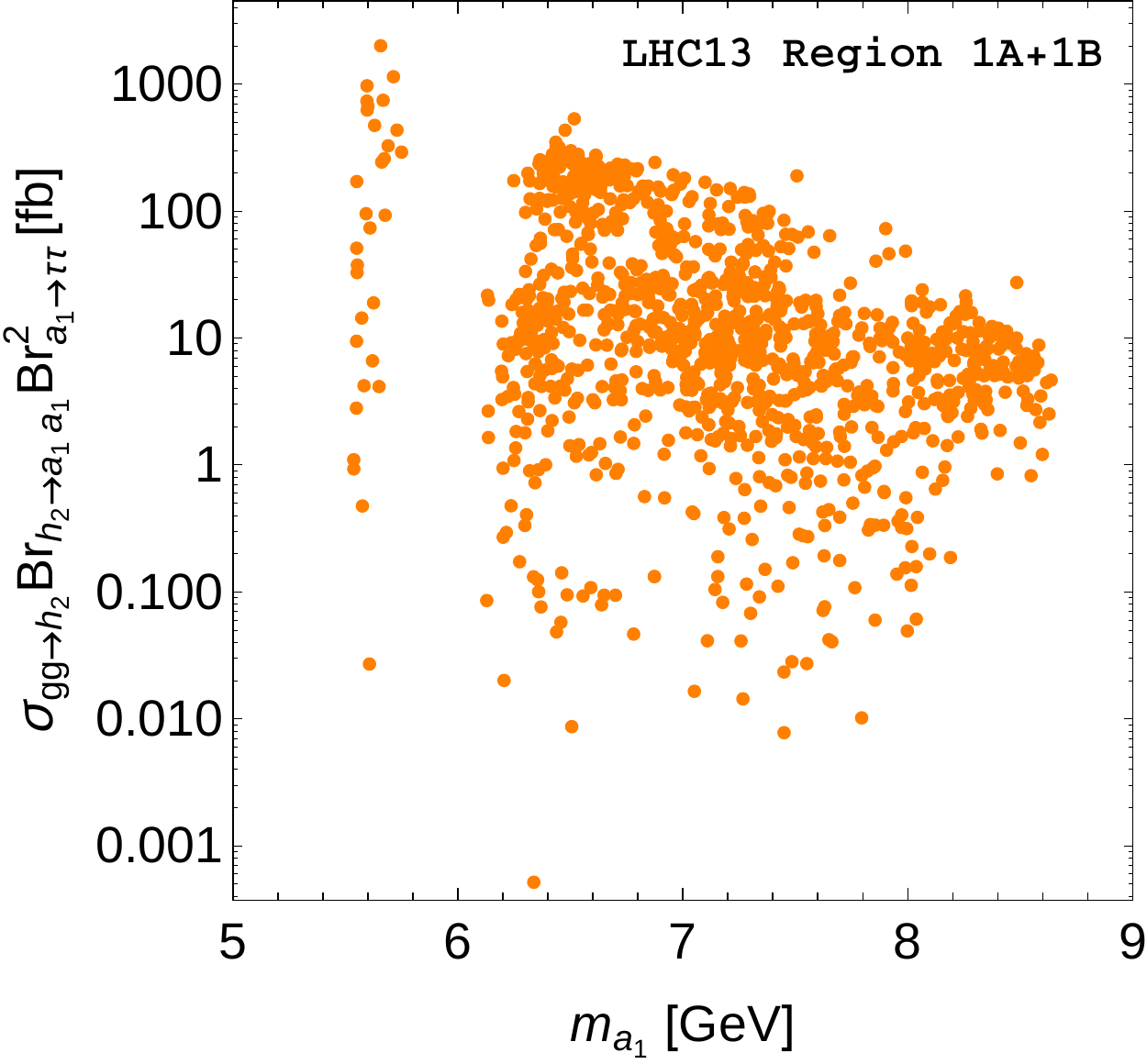}
\caption{$\sigma(gg\to h_2)/\sigma(gg\to h^{\rm SM}){\textrm{Br}}(h_2\to a_1 a_1){\textrm{Br}}^2(a_1\to \tau\tau)$ for the LHC Run-1 (left) and $\sigma(gg\to h_2){\textrm{Br}}(h_2\to a_1 a_1){\textrm{Br}}^2(a_1\to \tau\tau)$ for the LHC Run-2 (right) for region 1A and 1B. The horizontal line in the left plot is the limit set by ATLAS~\cite{Aad:2015oqa}.}
\label{fig:1AB_ttmumu}
\end{figure}

The leptonic channels have already been investigated at LHC Run-1. The ATLAS collaboration has performed a search for a CP even Higgs boson produced through gluon fusion and decaying into a pair of pseudoscalars, and has set a limit on the production cross section times branching ratio into a $4\tau$ final state normalized to the SM Higgs boson cross section~\cite{Aad:2015oqa}. This limit can be easily related to the $2\mu2\tau$ final state under the assumption that $Br(a_1\to\mu\mu)/Br(a_1\to\tau\tau)=m_\mu^2/(m_\tau^2\sqrt{1-(2m_\tau/m_{a_1}})^2)$. We show in Fig.~\ref{fig:1AB_ttmumu} (left panel) the quantity $\sigma(gg\to h_2)/\sigma(gg\to h^{\rm SM}){\textrm{Br}}(h_2\to a_1 a_1){\textrm{Br}}^2(a_1\to \tau\tau)$ for region 1A and 1B, together with the limit set by the ATLAS collaboration. The rescaled production cross section has been computed using the reduced $ggh_2$ coupling provided by \verb#NMSSMTools#, which is given with respect to a SM Higgs boson of the same mass. Given the small size of the deviations of the $h_2$ couplings with respect to the SM Higgs ($g_{ggh_2}/g^{\rm SM}_{ggh_2}\sim 1)$, and assuming $\sigma(gg\to h_2)_{\textrm SM}\sim 20$~pb for $m_h^{\rm SM}\sim 125$~GeV, this analysis sets a limit on the inclusive cross section times branching ratio of $\sim 2$~pb for $m_{a_1}=$ 5--10~GeV. While this analysis does not constrain at the moment region 1 of the nMSSM, prospects for LHC Run-2 are quite exciting, given an inclusive cross section that can reach the $\sim$~pb level in the $a_1$ considered mass range (right panel). A recently published CMS analysis~\cite{Khachatryan:2015wka} also sets limits on the same production process, decaying however into a $4\mu$ final state. In the case where the only source of signal events is given by $h_2\to a_1 a_1\to 4\mu$, the limit is set to 1~fb for $m_{a_1} = 3.55$~GeV, which is well above the maximum rate obtained in the nMSSM~($\sim 0.01$~fb).

As mentioned, other interesting possibilities are the decays of the SM like Higgs boson into a pair of lighter CP even Higgs states ($h_2 \to h_1h_1$), which then decay into a $2b2\tau$ final state for which rates could reach $\mathcal{O}(250~{\rm fb})$ at the 13~TeV LHC. The decay into a 4$b$ final state has a larger rate, $\mathcal{O}(1~{\rm pb})$, but suffers from a higher QCD background~\cite{Bomark:2014gya}. While our discussion has been carried out assuming dominant gluon fusion production of the SM like Higgs boson, the contribution of different production modes (vector boson fusion or Higgs Strahlung) can improve the signal over background ratio, despite smaller production cross sections. Finally, direct production of a light pseudoscalar $a_1$, either via gluon fusion or via $b\bar b$ associate production, can provide other possible signatures for the search of a light pseudoscalar~\cite{Bomark:2014gya}.

In region 1 the prospects for DM DD are quite pessimistic, due to the lightness of the LSP which has a mass below the lower limit that can be tested by the XENON1T experiment~\cite{Aprile:2012zx}. The value of the (rescaled) spin independent DD cross section, being at most $\sim 10^{-10}$~pb, is also problematic. This is about 1 order of magnitude below the neutrino coherent scattering background for a WIMP mass of $\sim 5$~GeV~\cite{Ruppin:2014bra}.

We conclude this Section proposing in Tab.~\ref{tab:1AB_BMP} three benchmark points in region 1A, with the Higgs sector signatures relevant for the 13~TeV run of the LHC discussed above. Tab.~\ref{tab:1AB_BMP} reports also the relevant masses, cross sections (obtained with \sushi~\verb#v1.5.0#~\cite{Harlander:2012pb,Liebler:2015bka} at NNLO) and branching ratios for the various benchmark points. These scenarios are amenable for a deeper phenomenological investigation, both from the theoretical and experimental side. The benchmark points BMP1A-I and BMP1A-II are in a configuration where $m_{h_2}>2 m_{h_1}$, and maximise Br$(h_2\to a_1 a_1)$ and Br$(h_2\to h_1 h_1)$ respectively, with the possibility of giving rise to the aforementioned multi-$\tau$ or multi-$b$ final states, although the branching ratios are only $\sim$4\% for 4$b$ and $\sim$1\% for $2b2\tau$. We also propose a third scenario, BMP1A-III, where the $h_2\to h_1 h_1$ channel is kinematically closed, and the decay rate of the SM like Higgs in two light pseudoscalars is below $10^{-3}$. The decay pattern of $h_1$ and $a_1$ in this configuration are similar to the ones of BMP1A-I and BMP1A-II, but clearly these states now need to be produced directly. We do not propose benchmark points for region 1B for two reasons. First, the characteristics of the Higgs sector are quite similar to the ones of region 1A when the decay pattern $h_2\to h_1 h_1$ is closed (recall in fact that in region 1B the $h_1$ mass is between 70 and 90~GeV). Secondly, and more importantly, the lighter gluino in this region is likely to be tested with early data from LHC Run-2 through conventional search channels.

In summary, the Run-2 of the LHC at higher energy and luminosity will further probe region 1 through stop, slepton and gluino searches, while new Higgs decay channels involving light pseudoscalar and/or scalar bosons can provide characteristic signatures of an extended Higgs sector. Moreover, peculiar signatures from stop decays could also characterise the light singlino scenario.

\begin{table}
\begin{center}
\begin{tabular}{| l | c |c | c |}
\hline
 & BMP1A-I & BMP1A-II & BMP1A-III \\
\hline
\hline
$\tb$ & 7.8 & 8.32 & 9.48 \\
\hline
$\l$ & 0.372 & 0.4 & 0.509 \\
\hline
$\mu$ & 265 & 290 & 375 \\ 
\hline
$m_0$ & 0 & 0 & 505 \\
\hline
$ M_{1/2}$ & 766 & 790 & 888 \\ 
\hline
$A_0$ & -1146 & -1050 & 9 \\
\hline
$A_\l$ & 2286 & 2700 & 5715 \\
\hline
$\xi_F$ & 28 & 48 & 79 \\
\hline
$\xi_S$ & -5.3$\cdot10^4$ & -4.6$\cdot10^4$ & 279 \\
\hline
\hline
Masses & $m_{h_1}=37.0$\ $m_{a_1}=6.8$ & $m_{h_1}=43.6$\ $m_{a_1}=6.8$ & $m_{h_1}=64.7$\ $m_{a_1}=8.0$ \\
\hline 
$\sigma^{13~\rm TeV}$ [pb] & $\sigma(gg\to h_2)=41.5$ & $\sigma(gg\to h_2)=42.2$ & $\sigma(gg\to h_2)=42.4$ \\
 & $\sigma(gg\to h_1)=13.0$ & $\sigma(gg\to h_1)=1.8$ & $\sigma(gg\to h_1)=1.5$\\
 & $\sigma(gg\to a_1)=242.8$ & $\sigma(gg\to a_1)=236.5$ & $\sigma(gg\to a_1)=244.5$ \\
\hline
Br($h_2$) & Br($h_2\to a_1a_1$)=8\% & Br($h_2\to h_1h_1$)=10\% & \\
\hline
Br($h_1$) & Br($h_1\to b\bar b$)=85\% & Br($h_1\to b\bar b$)=65\% & Br($h_1\to b\bar b$)=58\%\\
 & Br($h_1\to \tau\tau)=$7\% & Br($h_1\to a_1a_1$)=28\% & Br($h_1\to a_1a_1)=$33\% \\
 & Br($h_1\to \tilde\chi^0_1\tilde\chi^0_1$)=7\% & Br($h_1\to \tau\tau$)=6\% & Br($h_1\to \tau\tau$)=5\% \\
\hline 
Br($a_1$) &Br($a_1\to \tilde\chi^0_1\tilde\chi^0_1$)=73\% & Br($a_1\to \tilde\chi^0_1\tilde\chi^0_1$)=73\% & Br($a_1\to \tilde\chi^0_1\tilde\chi^0_1$)=78\% \\
 & Br($a_1\to \tau\tau$)=25\% & Br($a_1\to \tau\tau$)=25\% & Br($a_1\to \tau\tau$)=20\% \\
\hline 
\end{tabular}
\caption{Benchmark points choices for region 1A. Dimensionful parameter are expressed in GeV, except $\xi_F$ and $\xi_S$ which are in GeV$^2$ and GeV$^3$ respectively and $\sigma$, expressed in pb. We indicate in the table the relevant mass spectrum, cross sections and branching ratios.}
\label{tab:1AB_BMP}
\end{center}
\end{table}

\subsection{Region 2, $45 <m_{\tilde\chi^0_1}< 55$~GeV}

This region is characterised by a large $m_0 \sim 10^5$~GeV and a small $M_{1/2} \ltap 500$~GeV. The LSP is almost an equal admixture of higgsino and singlino, with a bino component $\lesssim$ 20\%. The DM relic density is always below $\Omega h^2<$ 0.01, see Fig.~\ref{fig:scan_res}. Due to the high value of $m_0$, all sfermions are heavy and decoupled and the only sub-TeV sparticles are EWinos and gluinos. The EWinos are rather mixed states, with a mass $\ltap 200$~GeV, while the gluino lies in the 800--1500~GeV mass range. The lightest CP even Higgs state, $h_1$, is SM like, and all the other Higgs states are heavier than 300~GeV, with $h_2$ and $a_1$ carrying a dominant singlet component if lighter than 600~GeV (otherwise $h_3$ and $a_2$ are the dominant singlet states).

To check the LHC Run-1 reach on the gluino sector, we have exploited the same technique used in Sec.~\ref{sec:1AB_constraints}, based on the recast of the CMS gluino search~\cite{CMS:2013ija} implemented in \ma5, and obtained a similar limit of $m_{\tilde g}\gtap 1100$~GeV.

\begin{figure}[htb]
\begin{center}
\includegraphics[width=0.46\textwidth]{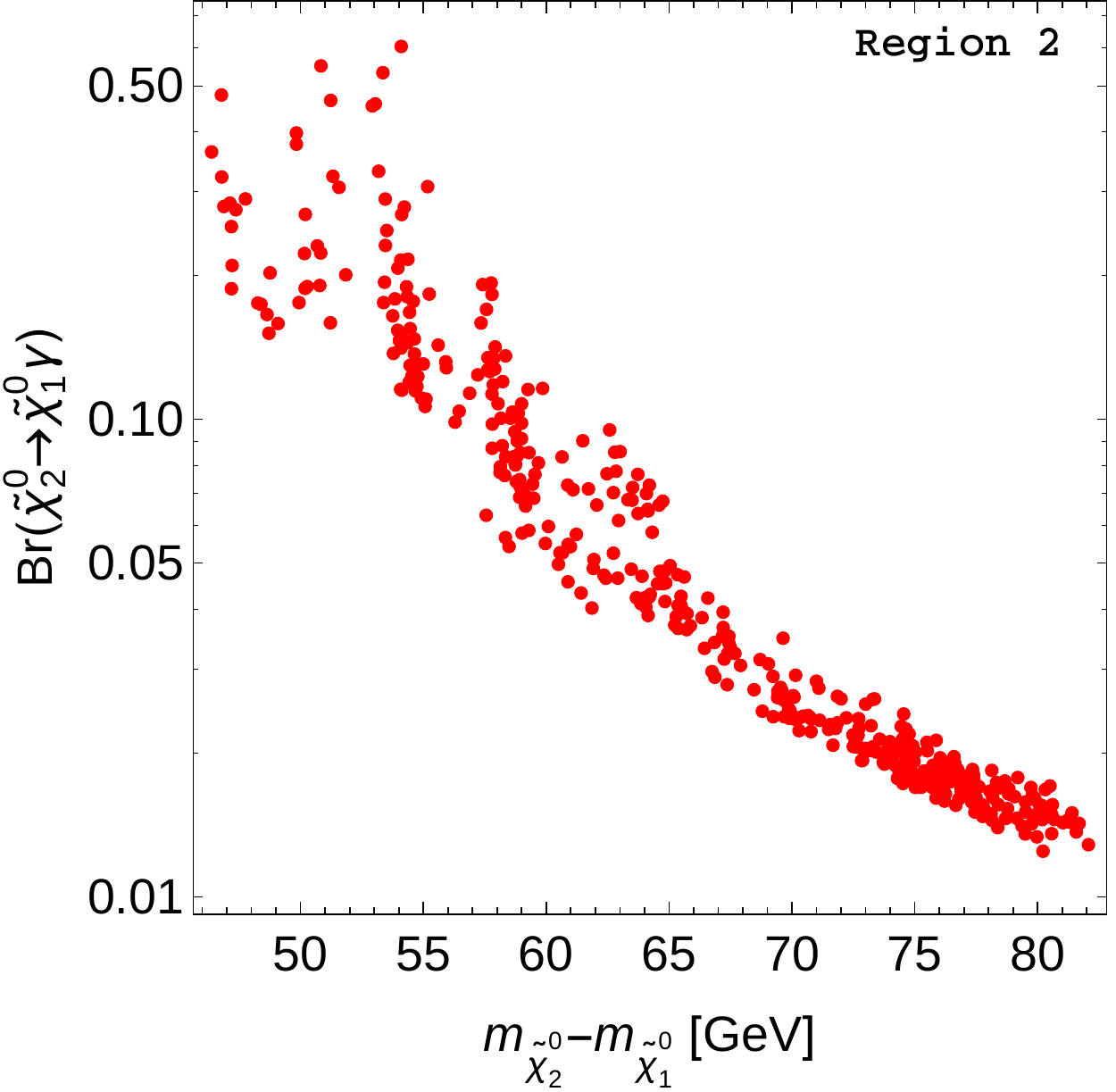} \hfill
\includegraphics[width=0.478\textwidth]{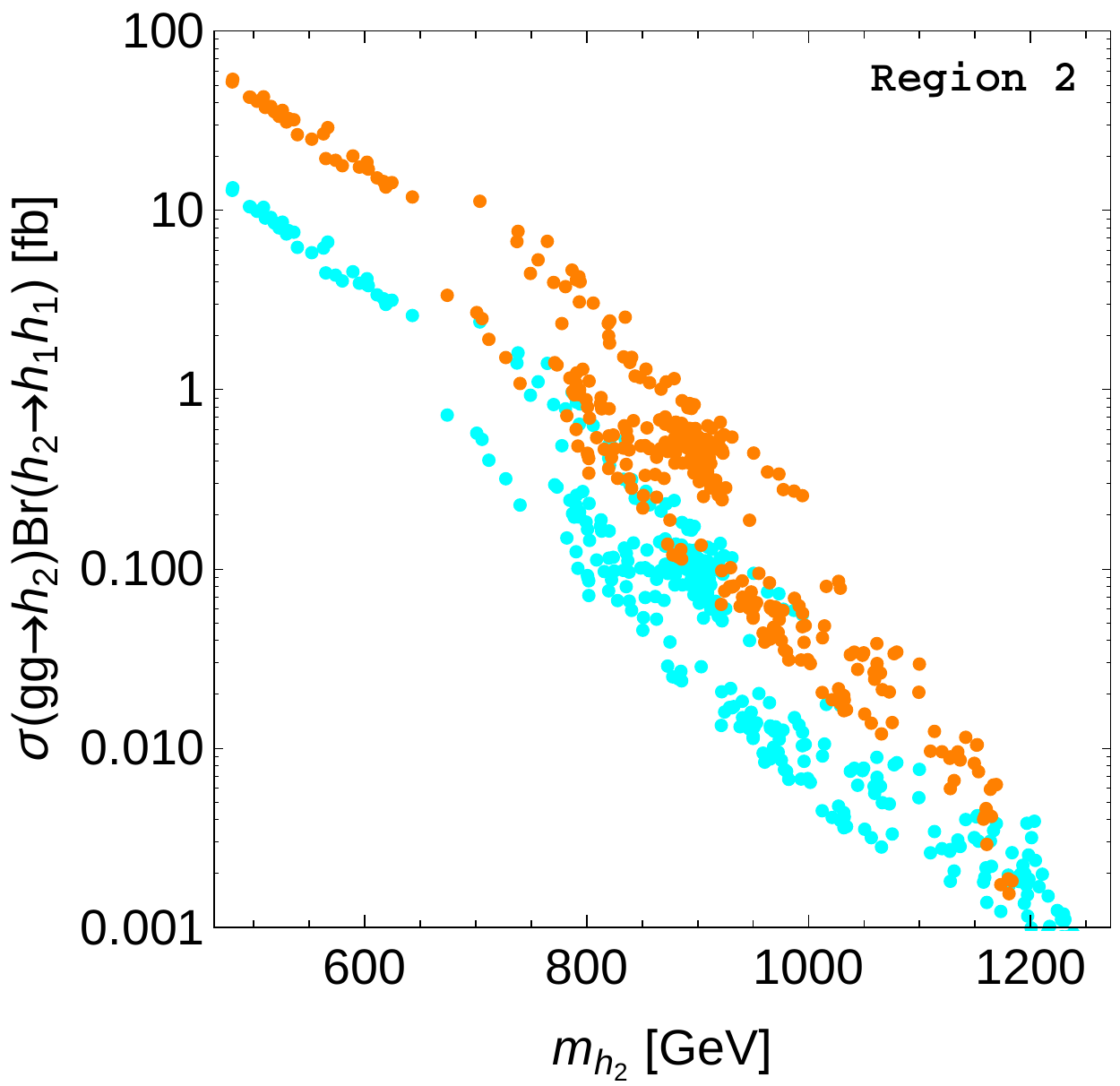}
\caption{Br$(\tilde\chi^0_2\to \gamma \tilde\chi^0_1)$ in function of $m_{\tilde\chi^0_2}-m_{\tilde\chi^0_1}$ (left) and $\sigma(gg\to h_2){\rm Br}(h_2\to h_1 h_1)$ in function of $m_{h_2}$ (right) for the 8 and 13~GeV LHC (cyan and orange points respectively). Both panels include only points in region 2 for which $m_{\tilde g}\gtap 1100$~GeV. The LSP mass is $\sim 45$--$55$~GeV.}
\label{fig:2_n2decay}
\end{center}
\end{figure}

Using \smodels~we did not find any additional constraint arising from the EWino sector. The most sensitive analysis, among the ones implemented in \smodels, is the CMS search for $\tilde\chi_2^0\tilde\chi^\pm_1$ production in $W^\pm Z + E_T^{\rm miss}$ final state with either on-shell or off-shell SM gauge bosons~\cite{Khachatryan:2014qwa}. However, in this region of the nMSSM parameter space, the masses of the EWinos are approximately $m_{\tilde\chi^\pm_1}\sim$103--120~GeV and $m_{\tilde\chi^0_2}\sim$~80--120~GeV. Their decays therefore occur via an off-shell $W$ or $Z$, falling close to the region where the CMS search looses sensitivity, $m_{\tilde\chi^0_2}=m_{\tilde\chi^0_1}+m_Z$. Another reason for this region not to be constrained by EWino searches is that there are other important decay channels for the neutralinos, which lead to a reduction of the branching into a Z boson. Notably, $\tilde\chi^0_2$ can decay into a $\gamma \tilde\chi^0_1$ final state, with a branching ratio up to $50\%$ for a mass splitting between $\tilde\chi^0_2$ and $\tilde\chi^0_1$ of $\sim 50$~GeV, as shown in Fig.~\ref{fig:2_n2decay} (left panel). This opens up the interesting possibility of looking at $\tilde\chi^\pm_1\chi_2^0$ production, for which the cross section can be close to $\mathcal{O}(1~{\rm pb})$ already at LHC Run-1, in either the $2j+\gamma+E_T^{\rm miss}$ or $l+\gamma+E_T^{\rm miss}$ final states. Interestingly, the photon might have a sizeable $p_T$, as the $\tilde\chi^0_2-\tilde\chi^0_1$ mass difference can be $\mathcal{O}$(50~GeV). However such a mass splitting does not allow for large $E_T^{\rm miss}$, therefore it might be necessary to trigger on an ISR jet. We leave this point for further investigation and we provide one benchmark point for this scenario in Tab.~\ref{tab:2_BM} (BMP2-I).

In this region the extra Higgs states $h_2$ and $a_1$ are either doublet or singlet like, depending on their masses, and almost degenerate. Only the states heavier than $\sim 450$~GeV survive LHC-Run 1 constraints, set by gluino searches. Possible decay patterns that have been searched for at the LHC are $a_1\to Z h_1$ and $h_2\to h_1h_1$~\cite{Khachatryan:2015lba,Aad:2015wra,Khachatryan:2014jya}, which do not however impose any constraint on this region of the parameter space. This is due to the small Br$(a_1\to Z h_1)<0.5$\% and to the fact that the mass of $h_2$ lies beyond the reach of the experimental analysis which is around 360~GeV.

We show in Fig.~\ref{fig:2_n2decay} (right panel) the rates $\sigma(gg\to h_2) {\rm Br}(h_2\to h_1 h_1)$ for the 8 and 13~TeV LHC (cyan and orange respectively), where the gluon fusion production cross sections have been obtained by rescaling the one provided by the LHC Higgs cross section working group~\cite{Heinemeyer:2013tqa} with the reduced $ggh_2$ coupling provided by \verb#NMSSMTools#. Recall that the current limit is $\sigma(gg\to h_2){\rm Br}(h_2\to h_1 h_1) \ltap 4$~pb for $m_{h_2}=360$~GeV, thus roughly two orders of magnitude above the predicted value. 

\begin{figure}[htb]
\includegraphics[width=0.46\textwidth]{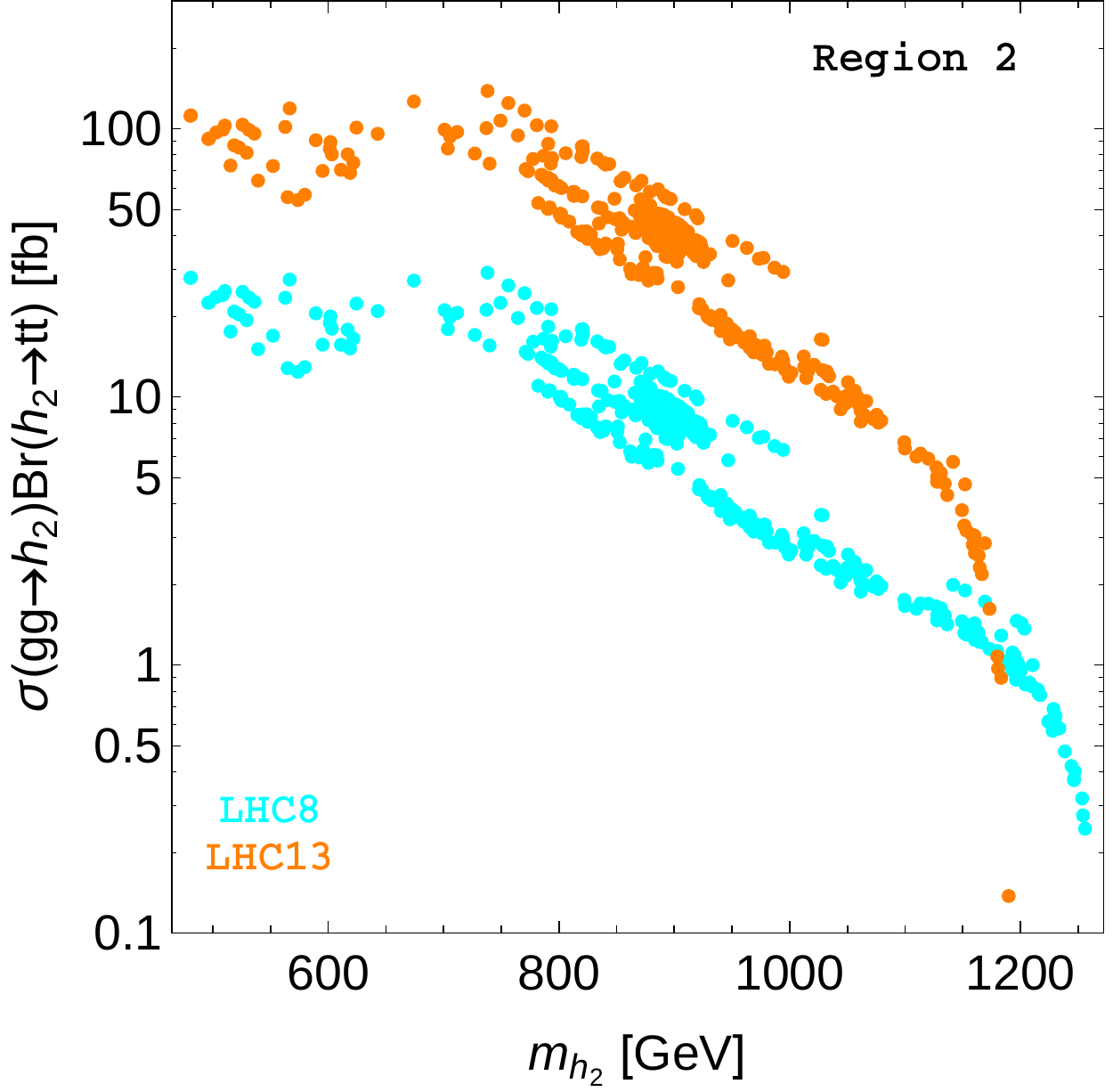}\hfill
\includegraphics[width=0.48\textwidth]{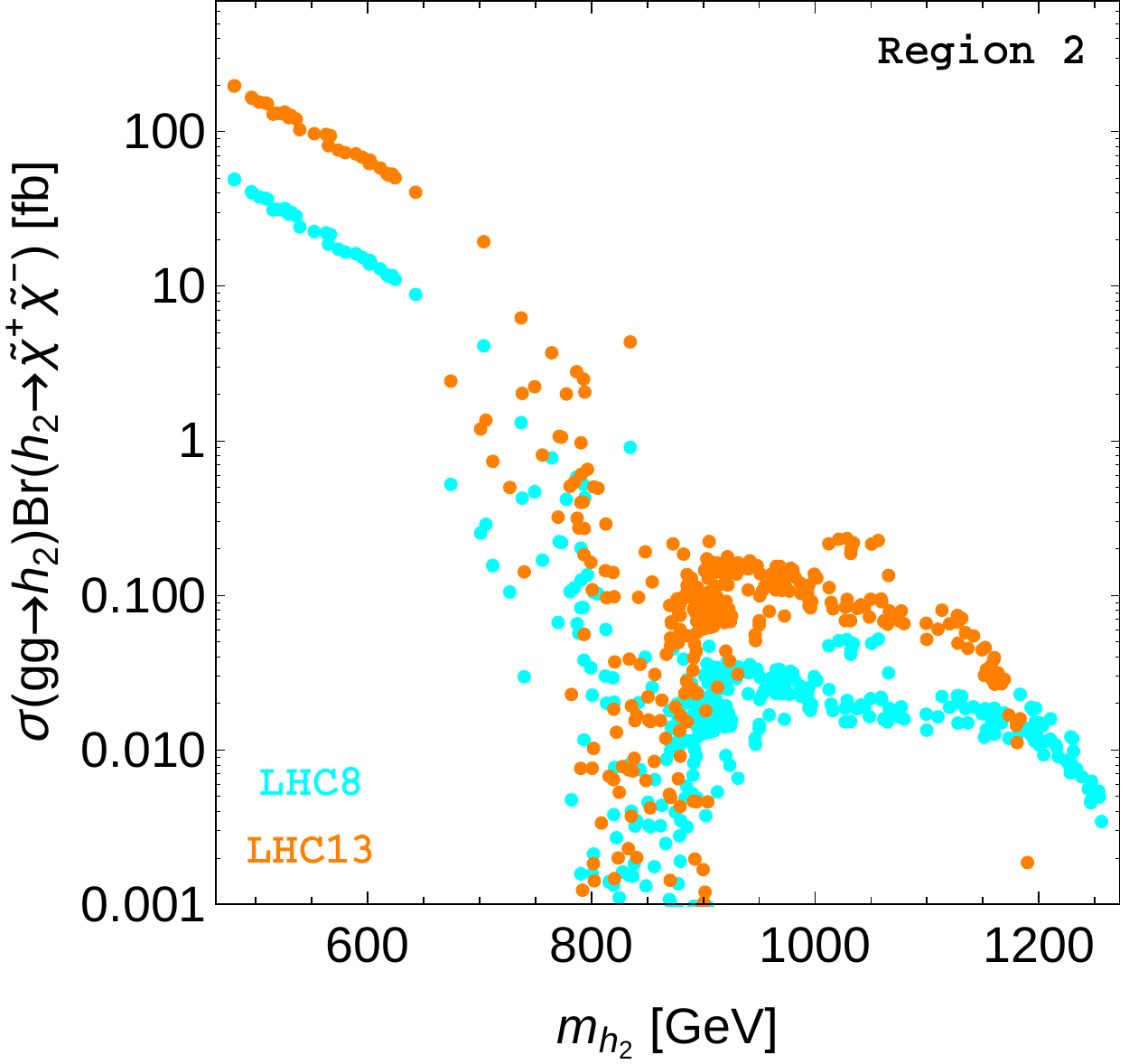}
\caption{$\sigma(gg\to h_2){\rm Br}(h_2\to t\bar t)$ (left) and $\sigma(gg\to h_2){\rm Br}(h_2 \to \tilde\chi^+_1\tilde\chi^-_1)$ (right) for the 8 and 13~TeV LHC (cyan and orange points respectively) as function of $m_{h_2}$.}
\label{fig:2_sigmah2ttchichi}
\end{figure}

Given that the mass of the $h_2$ is above the $t\bar t$ threshold, of particular interest is its decays into a top pair. The cross section times branching ratio can reach $\mathcal{O}(100~{\rm fb})$ for the 13~TeV LHC, as illustrated in Fig.~\ref{fig:2_sigmah2ttchichi} (left panel), and a similar value is predicted for $a_1$. This channel however deserves a deeper investigation, due to the well known possibility of dominant interference with the $t \bar t$ QCD background, for which dedicated cuts might need to be applied, see {\it e.g. }\cite{Barger:2006hm}. In the MSSM, an analysis has shown that the LHC could find evidence of a Higgs boson in this channel for masses up to 1~TeV and low values of $\tb$~\cite{Djouadi:2015jea}. The possibility of $h_2$ decaying into a pair of EWinos is also interesting, although it weakens the discovery potential in the $t\bar{t}$ mode. We show as an illustrative example the rates for the process $gg\to h_2\to \tilde\chi^+_1\tilde\chi^-_1$ in Fig.~\ref{fig:2_sigmah2ttchichi} (right panel) for the 8 and 13~TeV LHC. For the latter the cross section times branching can reach $\mathcal{O}(100~{\rm fb})$ for a $\sim 500$~GeV $h_2$, giving the possibility of a different search channel for $h_2$ and/or $a_1$. Clearly, for heavier Higgs states the rate drops rapidly. We provide in Tab.~\ref{tab:2_BM} a benchmark point (BMP2-II) that simultaneously maximises the two aforementioned rates. Note that for this point the branching ratio into neutralinos is similar to the one into charginos, while the pure invisible decay has a 6\% branching ratio. Rates for the lightest pseudoscalar $a_1$ are somewhat similar. Note that such large branching ratios in EWinos can also be found in the MSSM~\cite{Belanger:2015vwa}.

\begin{figure}[htb]
\includegraphics[width=0.490\textwidth]{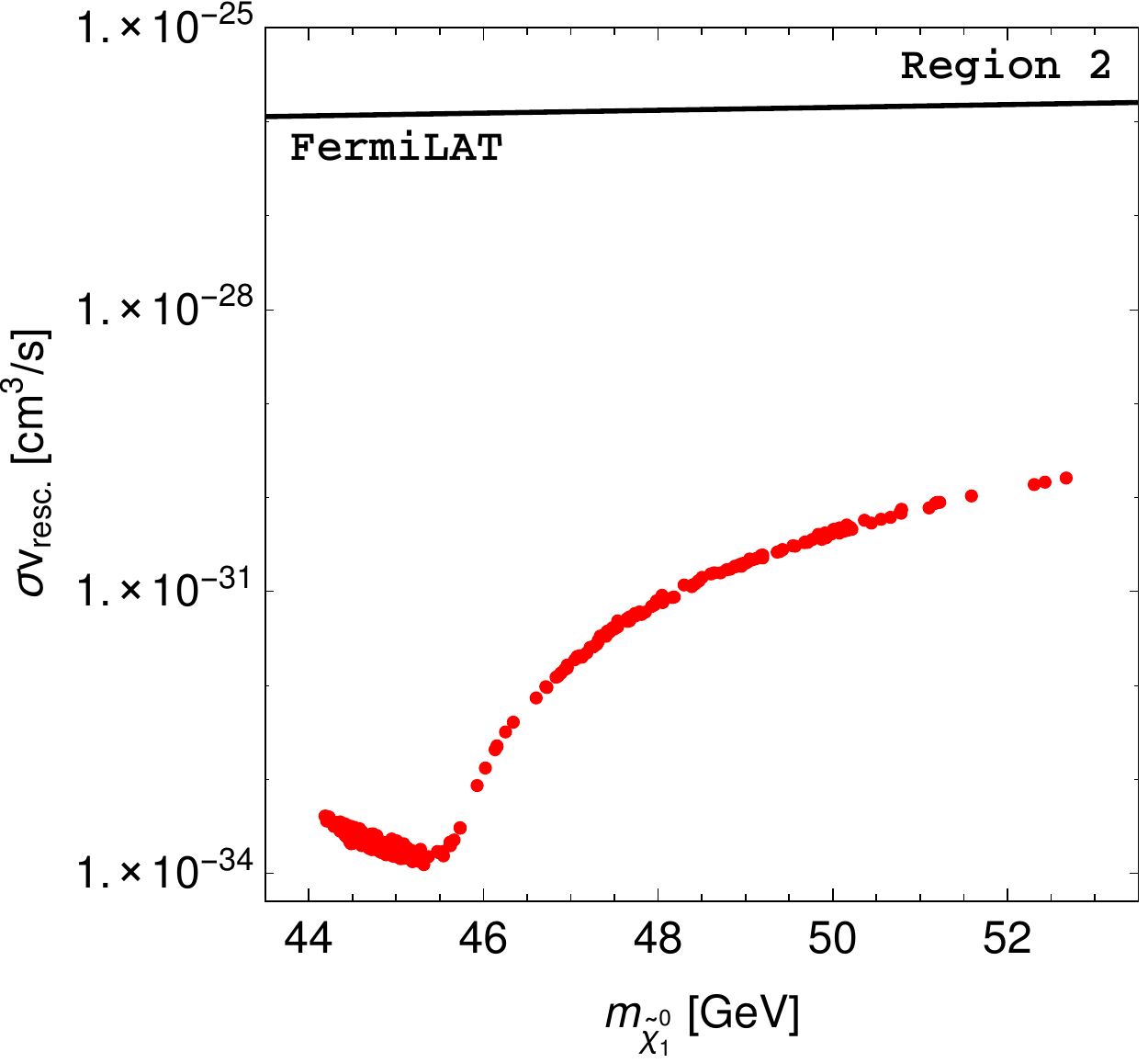}\hfill
\includegraphics[width=0.465\textwidth]{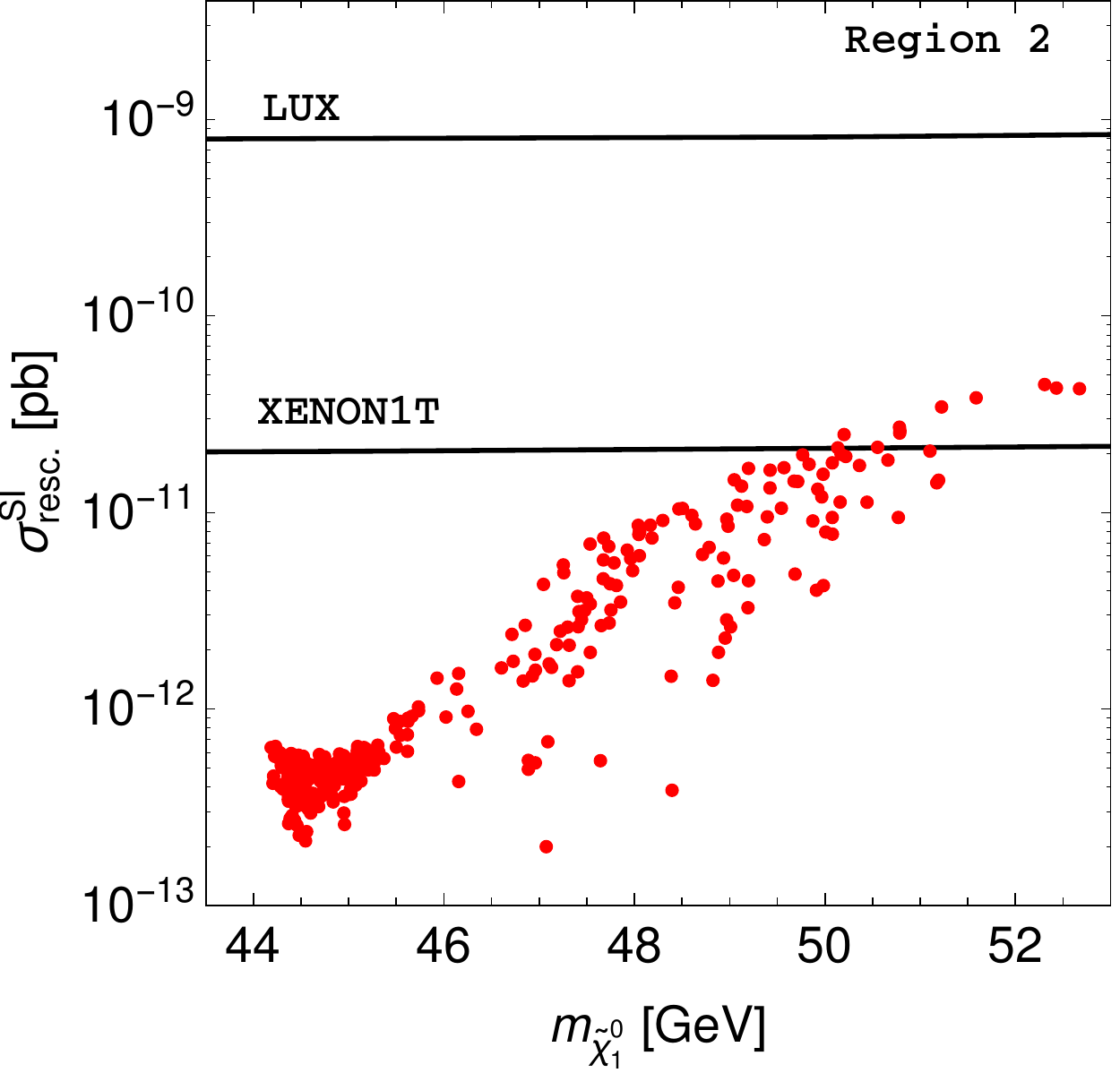}
\caption{ID detection rate into $b \bar b$ final state (left) and rescaled spin independent cross section for DD (right) in function of $m_{\tilde\chi^0_1}$.
The horizontal lines represent the current limit from FermiLAT (left) and LUX (right). Also shown in the right plot is the projection for XENON1T experiment. Only the points surviving the LHC Run-1 constraints are shown.}
\label{fig:2_DD_ID}
\end{figure}

Finally, unlike the first region, ID experiments do not set any additional constraints, since the rescaled rate into $b\bar b$ final state (the most relevant for the mass range of interest) is much below the current constraint from Fermi-LAT, as shown in Fig.~\ref{fig:2_DD_ID} (left panel). Prospects for DD measurements are once again quite pessimistic, as illustrated in the right panel of Fig.~\ref{fig:2_DD_ID}. The rescaled spin independent cross section for DD is more than one order of magnitude below the current limit set by the LUX experiment, and XENON1T will be able to marginally cover only a small region with $m_{\tilde\chi^0_1}> 50$~GeV. Such bleak DD prospects are typical of models where one relies on a resonance to achieve the correct relic density.

\begin{table}
\begin{center}
\begin{tabular}{| l | c | c | }
\hline
 & BMP2-I& BMP2-II \\
\hline
\hline
$\tb$ & 1.7 & 1.82 \\
\hline
$\l$ & 0.73 & 0.72 \\
\hline
$\mu$ & 130 & 126.0 \\ 
\hline
$m_0$ & 2.34$\cdot10^5$ & 5.3$\cdot10^4$ \\
\hline
$ M_{1/2}$ & 269 & 326 \\ 
\hline
$A_0$ & -5879 & -3180 \\
\hline
$A_\l$ & 0 & 0 \\
\hline
$\xi_F$ & 5.46$\cdot10^5$ & 1.52$\cdot10^5$ \\
\hline
$\xi_S$ & -301 & -2971 \\
\hline
\hline
Masses & $m_{\tilde\chi^\pm_1}=105$, $m_{\tilde\chi^0_2}=102$ & $m_{h_2}=478$, $m_{\tilde\chi^\pm_1}=104$ \\
\hline 
$\sigma$ [fb] &$\sigma^{13 (8)~\rm TeV}_{\tilde\chi^\pm_1\tilde\chi^0_2}=$ 1729 (860) & $\sigma^{13~\rm TeV}_{gg\to h_2}=$524 \\
\hline
Br &Br($\tilde\chi^0_2\to \tilde\chi^0_1\gamma$)=50\% & Br($h_2\to t\bar t$)=15\% \\
 && Br($h_2\to \tilde\chi^+_1\tilde\chi^-_1$)=29\% \\
 && Br($h_2\to \tilde\chi^0_{i\ne0} \tilde\chi^0_{j\ne 0})=$ 29\% \\
 && Br($h_2\to \tilde\chi^0_1 \tilde\chi^0_1)=$ 6\% \\
 && Br($\tilde\chi^\pm_1\to\tilde\chi^0_1 W^*$)=100\% \\
\hline 
\end{tabular}
\caption{Benchmark points choices for region 2. Dimensionful parameter are expressed in GeV, except $\xi_F$ and $\xi_S$ which are in GeV$^2$ and GeV$^3$ respectively and $\sigma$, expressed in fb. We indicate in the table the relevant mass spectrum, cross sections and branching ratios.}
\label{tab:2_BM}
\end{center}
\end{table}

\subsection{Region 3, $m_{\tilde\chi^0_1}\sim 65$~GeV}

The $\sim 65$~GeV LSP can appear in different regions of the $m_0$-$ M_{1/2}$ parameter space, in which it will have a different composition. We have identified the following three regions:
\begin{itemize}
 \item[-] Region 3A, large $m_0$ ($\sim~10^4$~GeV) and small $ M_{1/2}$ ($\ltap 200$~GeV). In this region the LSP is mainly bino and the DM relic density can be compatible with the value measured by Planck, see Fig.~\ref{fig:scan_res}.
 \item[-] Region 3B, $m_0\sim 1$~TeV and $ M_{1/2} \ltap 500$~GeV. In this region the LSP is a mixed higgsino and singlino state.
 \item[-] Region 3C, small $m_0$ ($\ltap 1$~TeV) and large $ M_{1/2}$ ($\sim$ 10$^4$~GeV). Here the LSP is also a higgsino/singlino admixture.
\end{itemize}

\subsubsection{Region 3A and 3B: bino or higgsino/singlino LSP with low $ M_{1/2}$}

In region 3A and 3B the relic density is kept to a small value by annihilation of the LSP through the Z boson or the SM like Higgs boson $h_1$. The very small higgsino component of the LSP in region 3A (around 2\%) is just sufficient to provide a DM relic density $\Omega h^2\approx 0.1$. The heavier CP even state, $h_2$, has a mass between 250 and 4000~GeV, almost degenerate with $a_1$, in both regions. Considering the values of $m_0$ and $ M_{1/2}$, the SUSY spectrum is characterised by light EWinos and light gluinos. Other sfermions are above the TeV scale and decoupled in region 3A while squarks, especially stops, can be below the TeV scale in region 3B.

In region 3A all the EWinos are lighter than 200~GeV, and the gluino is always lighter than 600~GeV. This region is therefore already completely ruled out by the LHC Run-1 searches for gluinos. Indeed, with the same procedure as in Sec.~\ref{sec:1AB_constraints}, we have checked that gluino masses $\lesssim 1100$~GeV are excluded. Recall that in a model with gaugino mass unification at the GUT scale, the gluino and bino masses are respectively $m_{\tilde g}\approx 3 M_{1/2}$ and $m_{\tilde B}\approx 1/6 M_{1/2}$, thus the upper limit on the (bino) LSP mass set by the upper limit on the singlino mass from Eq.~(\ref{eq:singlino}), entails an upper limit on the gluino mass. This has the important consequence of leaving just region 1, the one with the very light singlino LSP, as an nMSSM explanation for the entire relic abundance of the universe, together with a heavy enough gluino to pass LHC Run-1 constraints.

\begin{figure}[htb]
\begin{center}
\includegraphics[width=0.46\textwidth]{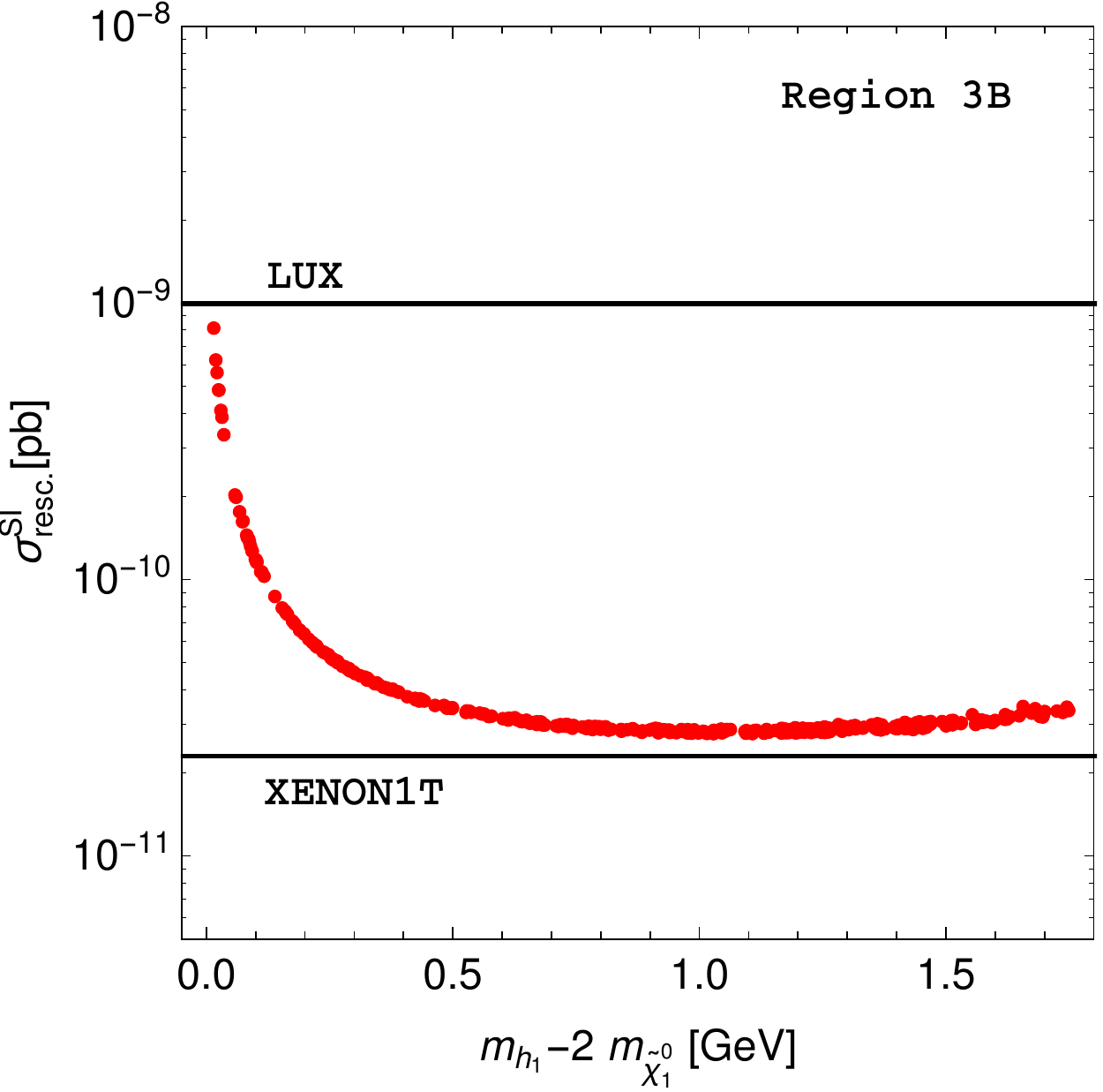}
\caption{Spin independent cross section for DD as function of $m_{h_1}-2m_{\tilde\chi^0_1}$, for all the points $m_{\tilde\chi^0_1}\approx 60 {\rm GeV}$
The horizontal lines represent the current limit from LUX and the projected limits for the XENON1T experiment.}
\label{fig:3A_DD}
\end{center}
\end{figure}

\begin{table}
\begin{center}
\begin{tabular}{| l | c |}
\hline
 & BMP3B-I\\
\hline
\hline
$\tb$ & 1.96 \\
\hline
$\l$ & 0.688 \\
\hline
$\mu$ & -105 \\ 
\hline
$m_0$ & 260 \\
\hline
$ M_{1/2}$ & 529 \\ 
\hline
$A_0$ & 3042 \\
\hline
$A_\l$ &1489 \\
\hline
$\xi_F$ & 1.66$\cdot10^6$ \\
\hline
$\xi_S$ & 3.82$\cdot10^8$ \\
\hline
\hline
Masses & $m_{h_2}=276$, $m_{\tilde\chi^\pm_1}=115$ \\
\hline 
$\sigma$ [fb] & $\sigma^{13~\rm TeV}_{gg\to h_2}=$ 180 \\
\hline
Br & Br($h_2\to\tilde\chi^+_1\tilde\chi^-_1$)=51\% \\
 & Br($\tilde\chi^\pm_1\to\tilde\chi^0_1 W^*$)=100\% \\
\hline
\end{tabular}
\caption{Benchmark points choices for region 3B. Dimensionful parameter are expressed in GeV, except $\xi_F$ and $\xi_S$ which are in GeV$^2$ and GeV$^3$ respectively and $\sigma$, expressed in fb. We indicate in the table the relevant mass spectrum, cross sections and branching ratios.}
\label{tab:3_BM}
\end{center}
\end{table}

The same bound on the gluino mass has also a strong impact in region 3B. Here the value of $ M_{1/2}$ is slightly higher than in region 3A, thus the gluino is somewhat heavier, $600 <m_{\tilde g}< 1200$~GeV. Nevertheless a large part of this region is already excluded by LHC Run-1. In addition, all EWinos except the heavier neutral and charged states, have a mass below 200~GeV, and are slightly constrained by direct EWinos searches, though the main constraint remains the one on the gluino mass. Moreover the lightest stop lies in the 200--500~GeV range because of the smaller value of $m_0$. In principle, for a 60~GeV LSP such a light stops could be excluded. However this is not the case here for two reasons. The first is that stops with a mass $\sim 250$~GeV are close to the kinematic edge $m_{\tilde t}-m_{\tilde\chi^0_1}=m_Z$ (see for example refs.~\cite{Chatrchyan:2013xna,Aad:2014kra}) where standard stop searches loose sensitivity, causing therefore a poor exclusion limit. The second is that for higher values of the stop mass, around 400~GeV, the same analyses loose sensitivity since the stop decays do not fulfill the simplified model assumptions. In particular, the branching ratios Br($\tilde t\to\tilde\chi^\pm_1 b)$ and Br($\tilde t\to\tilde\chi^0_1 t)$ are both suppressed. The reason is that the coupling of the stop to the singlino is small, hence other decay channels into heavier neutralinos are favoured despite the reduced phase space, causing therefore a reduction of the exclusion power of the experimental analyses. The two previous statements about the non exclusion of such light stops have been checked via the CMS search~\cite{Chatrchyan:2013xna} available in the \ma5 Public Analysis Database~\cite{MA5:CMS-SUS-13-011}. The reinterpretation of the LHC Run-1 results on gluino searches leave therefore just a small window of parameter space available, namely for $m_{\tilde g}\gtap 1100$~GeV. 

While extra Higgs searches such as $a_1\to Z h_1$ and $h_2\to h_1h_1$ do not set any constraint on this region of parameter space~\footnote{Note that the main search channel at low values of $\tb$ is in gauge bosons and that the constraints on the SM like Higgs imply a suppressed coupling of the heavy Higgs to gauge bosons, as in the MSSM. The suppression is even stronger for the singlet Higgs.}, exotic decays of heavy Higgs states in a pair of EWinos can have substantial rates. This leads again to the interesting possibility of searching for these states through their EWino decay channels. Cross sections of $\mathcal{O}$(100~fb) are expected for $gg\to h_2, h_2\to \tilde\chi^+_1\tilde\chi^-_1$ at the 13~TeV LHC, with a $h_2$ $\sim 300$~GeV. We provide one such benchmark in Tab.~\ref{tab:3_BM} (BMP3-I). Note that despite of a $h_2$ mass below 300~GeV, this point is safely below the recent limits of CMS in the $WW$ channel~\cite{Aad:2015agg}.

As concerns ID measurements, the relevant rates are well below the limits set by the Fermi-LAT experiment. Prospects for future underground experiments for DD are quite interesting.
The SI cross section lies just below the limits set by the LUX experiment and is roughly constant in this region.  It is determined essentially by the $h_1$ coupling to the LSP which does not vary much in this region. The rescaled cross section is however suppressed by up to almost  two orders of magnitude, since the thermally averaged DM annihilation cross section is  strongly enhanced when $m_{h_1}-2 m_{\tilde\chi_1^0}$ is  $\sim 1 $ GeV, see Fig.~\ref{fig:3A_DD}.
Nevertheless, as shown in this figure, XENON1T will be able to probe all this region of parameter space, offering a valid complement to collider based analyses.

\subsubsection{Region 3C: higgsino/singlino LSP with high $ M_{1/2}$}

This region at high $ M_{1/2}$ and low $m_0$ is characterised by light sleptons and light higgsinos and singlino, while the bino and wino states are heavy and decoupled. Squarks and gluinos are also heavy and decoupled, and the Higgs states $h_2$ and $a_1$ have a mass greater than 600~GeV. Note that even though $m_0$ is small at the GUT scale, the squarks are heavy because the renormalisation group equations that are used to derive their mass at the SUSY scale receive important contributions from $ M_{1/2}$ which is large. 

Despite the light EWinos, $m_{\tilde\chi^0_2,\tilde\chi^\pm_1}\sim 150$~GeV, sleptons, $m_{e_R/\mu_R}\sim 200$~GeV, and $\tilde \tau_1$ as light 100~GeV, no constraints are obtained by the LHC Run-1 SUSY searches. First and second generation right handed charged sleptons with a mass 100--300~GeV for a $\sim 65$~GeV LSP lie on the edge of the exclusion set by the ATLAS slepton search with $2l+E_T^{\rm miss}$ final state~\cite{Aad:2014vma}. The limit set by ATLAS on the $\tilde \tau^+\tilde \tau^-$ production cross section varies between $0.76$~pb for $m_{\tilde \tau} = 100$~GeV and $0.02$~pb for $m_{\tilde \tau}= 300$~GeV for a 60~GeV LSP~\cite{Aad:2014yka}. This limit is roughly one order of magnitude larger than the predicted values, therefore making this search not sensitive to this scenario.

\begin{figure}[htb]
\begin{center}
\includegraphics[width=0.46\textwidth]{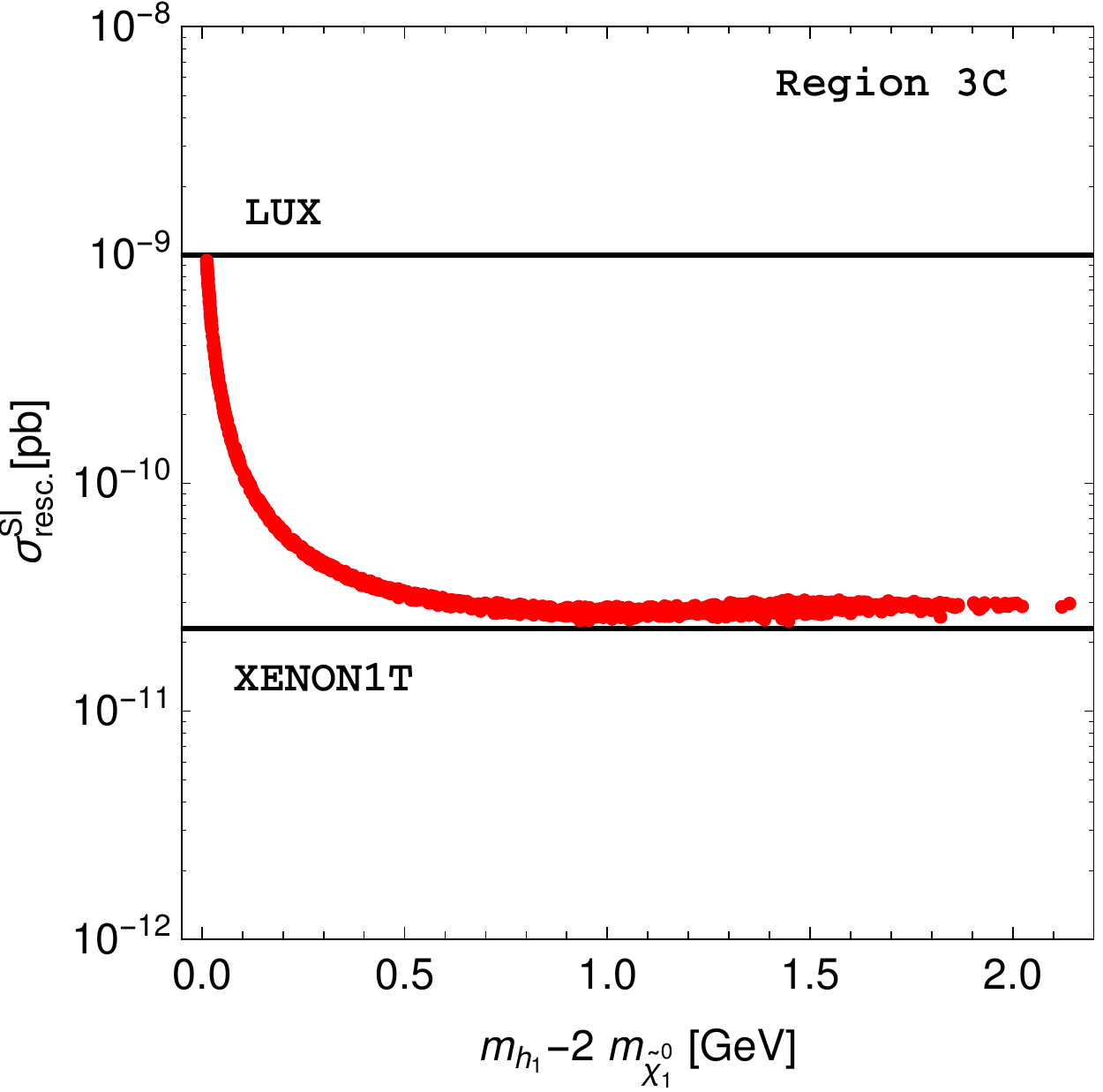}
\caption{
Spin independent cross section for DD as function of  $m_{h_1}-2m_{\tilde\chi^0_1}$.
The horizontal lines represent the current limit from LUX and the projected limits for the XENON1T experiment.}
\label{fig:stau_xs}
\end{center}
\end{figure}

Since the EWinos are mostly higgsino or singlino, their production cross sections are small. Moreover, the $\tilde\chi^0_{1,2}$ and $\tilde\chi^+_1$ masses lie close to the kinematic edge for off-shell decay where standard searches loose sensitivity, making it therefore easy to escape the LHC limits. We expect however that upgrade of the standard simplified model searches will be able to cover this region of parameter space with early LHC Run-2 data given the standard decay modes of these particles, and the fact that in this region \smodels~gives a ratio between the theoretical prediction and the experimental measurement which reaches the value of 0.7, {\it i.e.} not too far from the value of 1, for which we claim the point to be excluded.
Finally, as above, ton-scale underground experiments for DD, such as XENON1T, will be able to fully cover this region of parameter space, see Fig.~\ref{fig:stau_xs}, while no constraints are actually enforced by ID measurements.

\section{Conclusions}
\label{sec:sec5}

In this paper we have explored the parameter space of the nMSSM with input parameters defined at the GUT scale, and identified three broad regions for the LSP mass which satisfy theoretical, astrophysical, cosmological and collider constraints. In particular we have considered collider constraints on the Higgs boson and sparticles, as well as flavour constraints available in \nmssmtools, upper limit on DM DD from LUX and relic abundance of DM as measured by Planck. The nMSSM is characterised by the presence of a singlino with a mass below 75~GeV, since only mixing with higgsinos contribute to its, otherwise vanishing, mass. In only two of the three regions the lightest neutralino LSP can account completely for the DM abundance. The first features a light singlino LSP ($m_{\tilde\chi^0_1}\sim 5$~GeV) the second a heavier one ($m_{\tilde\chi^0_1}\sim 65$~GeV), mostly bino. In all the other allowed regions the computed relic density lies below the value measured by Planck.

A closer scrutiny at the constraints arising from SUSY searches at the LHC, implemented via two public tools for analyses recast, \smodels\ and \ma5, showed that some of these regions (or sub-regions) were actually completely exclude by results of LHC Run-1, notably when they featured a gluino below the TeV scale. In particular the only valid region not requiring an additional DM component is the one with a light singlino.

We have investigated how the model could be probed at the LHC Run-2 or through non collider DM searches. In particular, the region with a light singlino is characterised by the presence of a light pseudoscalar with a mass which is twice the one of the singlino, and a scalar Higgs with a mass below that of the recently discovered scalar. Hence, searches for these extra light states could lead to a distinctive signal of the nMSSM. Relevant signatures include the decay of the 125~GeV Higgs into light scalars or pseudoscalars, leading for example to $4b,2b/2\tau$ or $4\tau$ final states, or the direct production of one of the light scalar or pseudoscalar. Some of these signatures are under investigation at LHC, since the possibility of a light CP odd Higgs is one of the hallmark signature of generic NMSSM, and searches for light Higgs bosons are a priority in the LHC Run-2 program.

In other regions where the lightest Higgs is the already discovered boson at 125~GeV, the model can be probed via searches for the heavy Higgs states. One characteristic of the model, as opposed to the MSSM but shared with the generic NMSSM, is that values of $\tb\approx 1$--$2$ can be compatible with all the present constraints. Thus, searches for heavy Higgs states decaying into top pairs will provide further probes of the model. Another characteristic feature is the possibility of a heavy Higgs state decaying with a high rate into a pair of charginos or neutralinos. In particular we have found large rates even with a Higgs mass below 300~GeV. While a large branching ratio into charginos for the heavy Higgs state can occur in the MSSM, the Higgs mass is typically much higher~\cite{Belanger:2015vwa}, hence this process suffers from suppression due to the somewhat smaller cross section.

In the nMSSM the decays of sparticles have also peculiar features, thus searches for SUSY in Run-2 of the LHC could provide not only a validation of the theoretical SUSY framework, but also a sign of its non-minimality. The most distinctive feature occurs once again in the case of a light singlino LSP. Since the singlino is very weakly coupled to the sfermions, the decay channels involving the singlino in the final state typically have small partial widths. Therefore decays of sfermions, and in particular squarks, proceed preferably through the heavy neutralinos rather than directly in the LSP. Moreover the heavier neutralinos and charginos can decay into the LSP and a gauge boson (always permitted by phase space) and into a LSP and a light scalar or pseudoscalar. When the singlino LSP is not so light, we have found another unusual signature corresponding to $\tilde\chi_2^0$, which can decay with a substantial rate into a $\tilde\chi^0_1\gamma$ final state even when there is a considerable mass difference between these two states. Of course conventional SUSY searches at LHC Run-2 will probe further the allowed parameter space of the model, notably through searches for squarks and gluinos but also EWinos and sleptons.

 We have also pointed out that DD searches are complementary to collider searches, and that in particular a ton-scale detector could completely probe the remaining allowed region with a LSP around 65~GeV, while large signals in ID are not expected, except in very specific kinematic configurations. A thorough investigation of these new signatures at the LHC, which lies beyond the scope of this work, is required to assess the potential of the LHC to discover and/or probe the nMSSM.

\section*{Acknowledgments}

DB thanks Ursula Laa for useful discussions regarding the use of \smodels. This work was supported in part by the LIA-TCAP of CNRS, by the French ANR, Project DMAstro-LHC, ANR-12-BS05-0006, by the {\it Investissements d'avenir}, Labex ENIGMASS. The work of AP was also supported by the Russian foundation for Basic Research, grant RFBR-15-52-16021-CNRS-a. The authors acknowledge the support of France Grilles for providing cloud computing resources on the French National Grid Infrastructure.

\bibliographystyle{JHEP}
\bibliography{nMSSM}

\providecommand{\href}[2]{#2}\begingroup\raggedright\begin{thebibliography}{100}

\bibitem{Aad:2012tfa}
{\bf ATLAS} Collaboration, G.~Aad {\em et.~al.}, {\it {Observation of a new
  particle in the search for the Standard Model Higgs boson with the ATLAS
  detector at the LHC}},  {\em Phys.Lett.} {\bf B716} (2012) 1--29,
  [\href{http://xxx.lanl.gov/abs/1207.7214}{{\tt 1207.7214}}].

\bibitem{Chatrchyan:2012ufa}
{\bf CMS} Collaboration, S.~Chatrchyan {\em et.~al.}, {\it {Observation of a
  new boson at a mass of 125 GeV with the CMS experiment at the LHC}},  {\em
  Phys.Lett.} {\bf B716} (2012) 30--61,
  [\href{http://xxx.lanl.gov/abs/1207.7235}{{\tt 1207.7235}}].

\bibitem{Barbieri:1987fn}
R.~Barbieri and G.~F. Giudice, {\it {Upper Bounds on Supersymmetric Particle
  Masses}},  {\em Nucl. Phys.} {\bf B306} (1988) 63.

\bibitem{Hall:2011aa}
L.~J. Hall, D.~Pinner, and J.~T. Ruderman, {\it {A Natural SUSY Higgs Near 126
  GeV}},  {\em JHEP} {\bf 04} (2012) 131,
  [\href{http://xxx.lanl.gov/abs/1112.2703}{{\tt 1112.2703}}].

\bibitem{Maniatis:2009re}
M.~Maniatis, {\it {The Next-to-Minimal Supersymmetric extension of the Standard
  Model reviewed}},  {\em Int. J. Mod. Phys.} {\bf A25} (2010) 3505--3602,
  [\href{http://xxx.lanl.gov/abs/0906.0777}{{\tt 0906.0777}}].

\bibitem{Ellwanger:2009dp}
U.~Ellwanger, C.~Hugonie, and A.~M. Teixeira, {\it {The Next-to-Minimal
  Supersymmetric Standard Model}},  {\em Phys.Rept.} {\bf 496} (2010) 1--77,
  [\href{http://xxx.lanl.gov/abs/0910.1785}{{\tt 0910.1785}}].

\bibitem{Ellwanger:2011mu}
U.~Ellwanger, G.~Espitalier-Noel, and C.~Hugonie, {\it {Naturalness and Fine
  Tuning in the NMSSM: Implications of Early LHC Results}},  {\em JHEP} {\bf
  1109} (2011) 105, [\href{http://xxx.lanl.gov/abs/1107.2472}{{\tt
  1107.2472}}].

\bibitem{Cao:2012fz}
J.-J. Cao, Z.-X. Heng, J.~M. Yang, Y.-M. Zhang, and J.-Y. Zhu, {\it {A SM-like
  Higgs near 125 GeV in low energy SUSY: a comparative study for MSSM and
  NMSSM}},  {\em JHEP} {\bf 03} (2012) 086,
  [\href{http://xxx.lanl.gov/abs/1202.5821}{{\tt 1202.5821}}].

\bibitem{Ellwanger:2012ke}
U.~Ellwanger and C.~Hugonie, {\it {Higgs bosons near 125 GeV in the NMSSM with
  constraints at the GUT scale}},  {\em Adv.High Energy Phys.} {\bf 2012}
  (2012) 625389, [\href{http://xxx.lanl.gov/abs/1203.5048}{{\tt 1203.5048}}].

\bibitem{Perelstein:2012qg}
M.~Perelstein and B.~Shakya, {\it {XENON100 implications for naturalness in the
  MSSM, NMSSM, and $\lambda$-supersymmetry model}},  {\em Phys. Rev.} {\bf D88}
  (2013), no.~7 075003, [\href{http://xxx.lanl.gov/abs/1208.0833}{{\tt
  1208.0833}}].

\bibitem{Agashe:2012zq}
K.~Agashe, Y.~Cui, and R.~Franceschini, {\it {Natural Islands for a 125 GeV
  Higgs in the scale-invariant NMSSM}},  {\em JHEP} {\bf 02} (2013) 031,
  [\href{http://xxx.lanl.gov/abs/1209.2115}{{\tt 1209.2115}}].

\bibitem{Gherghetta:2012gb}
T.~Gherghetta, B.~von Harling, A.~D. Medina, and M.~A. Schmidt, {\it {The
  Scale-Invariant NMSSM and the 126 GeV Higgs Boson}},  {\em JHEP} {\bf 02}
  (2013) 032, [\href{http://xxx.lanl.gov/abs/1212.5243}{{\tt 1212.5243}}].

\bibitem{Cheng:2013fma}
T.~Cheng, J.~Li, T.~Li, and Q.-S. Yan, {\it {Natural NMSSM confronting with the
  LHC7-8}},  {\em Phys. Rev.} {\bf D89} (2014), no.~1 015015,
  [\href{http://xxx.lanl.gov/abs/1304.3182}{{\tt 1304.3182}}].

\bibitem{Kim:2013uxa}
D.~Kim, P.~Athron, C.~Balazs, B.~Farmer, and E.~Hutchison, {\it {Bayesian
  naturalness of the CMSSM and CNMSSM}},  {\em Phys. Rev.} {\bf D90} (2014),
  no.~5 055008, [\href{http://xxx.lanl.gov/abs/1312.4150}{{\tt 1312.4150}}].

\bibitem{Fowlie:2014faa}
A.~Fowlie, {\it {Is the CNMSSM more credible than the CMSSM?}},  {\em Eur.
  Phys. J.} {\bf C74} (2014), no.~10 3105,
  [\href{http://xxx.lanl.gov/abs/1407.7534}{{\tt 1407.7534}}].

\bibitem{Kim:1983dt}
J.~E. Kim and H.~P. Nilles, {\it {The mu Problem and the Strong CP Problem}},
  {\em Phys. Lett.} {\bf B138} (1984) 150.

\bibitem{Munir:2013wka}
S.~Munir, L.~Roszkowski, and S.~Trojanowski, {\it {Simultaneous enhancement in
  $\gamma \gamma, b\bar{b}$ and $\tau^{+} \tau^{-}$ rates in the NMSSM with
  nearly degenerate scalar and pseudoscalar Higgs bosons}},  {\em Phys. Rev.}
  {\bf D88} (2013), no.~5 055017,
  [\href{http://xxx.lanl.gov/abs/1305.0591}{{\tt 1305.0591}}].

\bibitem{Belanger:2014roa}
G.~Belanger, V.~Bizouard, and G.~Chalons, {\it {Boosting Higgs boson decays
  into gamma and a Z in the NMSSM}},  {\em Phys. Rev.} {\bf D89} (2014), no.~9
  095023, [\href{http://xxx.lanl.gov/abs/1402.3522}{{\tt 1402.3522}}].

\bibitem{Ellwanger:2014hia}
U.~Ellwanger and A.~M. Teixeira, {\it {NMSSM with a singlino LSP: possible
  challenges for searches for supersymmetry at the LHC}},  {\em JHEP} {\bf
  1410} (2014) 113, [\href{http://xxx.lanl.gov/abs/1406.7221}{{\tt
  1406.7221}}].

\bibitem{Jeong:2014xaa}
K.~S. Jeong, Y.~Shoji, and M.~Yamaguchi, {\it {Higgs Mixing in the NMSSM and
  Light Higgsinos}},  {\em JHEP} {\bf 11} (2014) 148,
  [\href{http://xxx.lanl.gov/abs/1407.0955}{{\tt 1407.0955}}].

\bibitem{King:2014xwa}
S.~F. King, M.~Muhlleitner, R.~Nevzorov, and K.~Walz, {\it {Discovery Prospects
  for NMSSM Higgs Bosons at the High-Energy Large Hadron Collider}},  {\em
  Phys. Rev.} {\bf D90} (2014), no.~9 095014,
  [\href{http://xxx.lanl.gov/abs/1408.1120}{{\tt 1408.1120}}].

\bibitem{Ellwanger:2014hca}
U.~Ellwanger and A.~M. Teixeira, {\it {Excessive Higgs pair production with
  little MET from squarks and gluinos in the NMSSM}},  {\em JHEP} {\bf 04}
  (2015) 172, [\href{http://xxx.lanl.gov/abs/1412.6394}{{\tt 1412.6394}}].

\bibitem{Bomark:2015fga}
N.-E. Bomark, S.~Moretti, and L.~Roszkowski, {\it {Detection prospects of light
  NMSSM Higgs pseudoscalar via cascades of heavier scalars from vector boson
  fusion and Higgs-strahlung}},  \href{http://xxx.lanl.gov/abs/1503.04228}{{\tt
  1503.04228}}.

\bibitem{Chakraborty:2015xia}
A.~Chakraborty, D.~K. Ghosh, S.~Mondal, S.~Poddar, and D.~Sengupta, {\it
  {Probing the NMSSM via Higgs boson signatures from stop cascade decays at the
  LHC}},  {\em Phys. Rev.} {\bf D91} (2015) 115018,
  [\href{http://xxx.lanl.gov/abs/1503.07592}{{\tt 1503.07592}}].

\bibitem{Potter:2015wsa}
C.~T. Potter, {\it {Natural NMSSM with a Light Pseudoscalar Higgs and Singlino
  LSP}},  \href{http://xxx.lanl.gov/abs/1505.05554}{{\tt 1505.05554}}.

\bibitem{Das:2012rr}
D.~Das, U.~Ellwanger, and A.~M. Teixeira, {\it {Modified Signals for
  Supersymmetry in the NMSSM with a Singlino-like LSP}},  {\em JHEP} {\bf 1204}
  (2012) 067, [\href{http://xxx.lanl.gov/abs/1202.5244}{{\tt 1202.5244}}].

\bibitem{Belanger:2005kh}
G.~Belanger, F.~Boudjema, C.~Hugonie, A.~Pukhov, and A.~Semenov, {\it {Relic
  density of dark matter in the NMSSM}},  {\em JCAP} {\bf 0509} (2005) 001,
  [\href{http://xxx.lanl.gov/abs/hep-ph/0505142}{{\tt hep-ph/0505142}}].

\bibitem{Vasquez:2012hn}
D.~A. Vasquez, G.~Belanger, C.~Boehm, J.~Da~Silva, P.~Richardson, {\em
  et.~al.}, {\it {The 125 GeV Higgs in the NMSSM in light of LHC results and
  astrophysics constraints}},  {\em Phys.Rev.} {\bf D86} (2012) 035023,
  [\href{http://xxx.lanl.gov/abs/1203.3446}{{\tt 1203.3446}}].

\bibitem{Hugonie:2007vd}
C.~Hugonie, G.~Belanger, and A.~Pukhov, {\it {Dark matter in the constrained
  NMSSM}},  {\em JCAP} {\bf 0711} (2007) 009,
  [\href{http://xxx.lanl.gov/abs/0707.0628}{{\tt 0707.0628}}].

\bibitem{Belanger:2008nt}
G.~Belanger, C.~Hugonie, and A.~Pukhov, {\it {Precision measurements, dark
  matter direct detection and LHC Higgs searches in a constrained NMSSM}},
  {\em JCAP} {\bf 0901} (2009) 023,
  [\href{http://xxx.lanl.gov/abs/0811.3224}{{\tt 0811.3224}}].

\bibitem{Ellwanger:2014dfa}
U.~Ellwanger and C.~Hugonie, {\it {The semi-constrained NMSSM satisfying bounds
  from the LHC, LUX and Planck}},  {\em JHEP} {\bf 08} (2014) 046,
  [\href{http://xxx.lanl.gov/abs/1405.6647}{{\tt 1405.6647}}].

\bibitem{Vasquez:2010ru}
D.~A. Vasquez, G.~Belanger, C.~Boehm, A.~Pukhov, and J.~Silk, {\it {Can
  neutralinos in the MSSM and NMSSM scenarios still be light?}},  {\em
  Phys.Rev.} {\bf D82} (2010) 115027,
  [\href{http://xxx.lanl.gov/abs/1009.4380}{{\tt 1009.4380}}].

\bibitem{AlbornozVasquez:2011js}
D.~Albornoz~Vasquez, G.~Belanger, and C.~Boehm, {\it {Astrophysical limits on
  light NMSSM neutralinos}},  {\em Phys.Rev.} {\bf D84} (2011) 095008,
  [\href{http://xxx.lanl.gov/abs/1107.1614}{{\tt 1107.1614}}].

\bibitem{AlbornozVasquez:2012px}
D.~Albornoz~Vasquez, G.~Belanger, J.~Billard, and F.~Mayet, {\it {Probing
  neutralino dark matter in the MSSM \& the NMSSM with directional detection}},
   {\em Phys.Rev.} {\bf D85} (2012) 055023,
  [\href{http://xxx.lanl.gov/abs/1201.6150}{{\tt 1201.6150}}].

\bibitem{Panagiotakopoulos:1999ah}
C.~Panagiotakopoulos and K.~Tamvakis, {\it {New minimal extension of MSSM}},
  {\em Phys. Lett.} {\bf B469} (1999) 145--148,
  [\href{http://xxx.lanl.gov/abs/hep-ph/9908351}{{\tt hep-ph/9908351}}].

\bibitem{Panagiotakopoulos:2000wp}
C.~Panagiotakopoulos and A.~Pilaftsis, {\it {Higgs scalars in the minimal
  nonminimal supersymmetric standard model}},  {\em Phys.Rev.} {\bf D63} (2001)
  055003, [\href{http://xxx.lanl.gov/abs/hep-ph/0008268}{{\tt
  hep-ph/0008268}}].

\bibitem{Dedes:2000jp}
A.~Dedes, C.~Hugonie, S.~Moretti, and K.~Tamvakis, {\it {Phenomenology of a new
  minimal supersymmetric extension of the standard model}},  {\em Phys.Rev.}
  {\bf D63} (2001) 055009, [\href{http://xxx.lanl.gov/abs/hep-ph/0009125}{{\tt
  hep-ph/0009125}}].

\bibitem{Panagiotakopoulos:2001zy}
C.~Panagiotakopoulos and A.~Pilaftsis, {\it {Light charged Higgs boson and
  supersymmetry}},  {\em Phys. Lett.} {\bf B505} (2001) 184--190,
  [\href{http://xxx.lanl.gov/abs/hep-ph/0101266}{{\tt hep-ph/0101266}}].

\bibitem{Menon:2004wv}
A.~Menon, D.~E. Morrissey, and C.~E.~M. Wagner, {\it {Electroweak baryogenesis
  and dark matter in the nMSSM}},  {\em Phys. Rev.} {\bf D70} (2004) 035005,
  [\href{http://xxx.lanl.gov/abs/hep-ph/0404184}{{\tt hep-ph/0404184}}].

\bibitem{Barger:2005hb}
V.~Barger, P.~Langacker, and H.-S. Lee, {\it {Lightest neutralino in extensions
  of the MSSM}},  {\em Phys. Lett.} {\bf B630} (2005) 85--99,
  [\href{http://xxx.lanl.gov/abs/hep-ph/0508027}{{\tt hep-ph/0508027}}].

\bibitem{Barger:2006dh}
V.~Barger, P.~Langacker, H.-S. Lee, and G.~Shaughnessy, {\it {Higgs Sector in
  Extensions of the MSSM}},  {\em Phys. Rev.} {\bf D73} (2006) 115010,
  [\href{http://xxx.lanl.gov/abs/hep-ph/0603247}{{\tt hep-ph/0603247}}].

\bibitem{Huber:2006wf}
S.~J. Huber, T.~Konstandin, T.~Prokopec, and M.~G. Schmidt, {\it {Electroweak
  Phase Transition and Baryogenesis in the nMSSM}},  {\em Nucl. Phys.} {\bf
  B757} (2006) 172--196, [\href{http://xxx.lanl.gov/abs/hep-ph/0606298}{{\tt
  hep-ph/0606298}}].

\bibitem{Barger:2006kt}
V.~Barger, P.~Langacker, and G.~Shaughnessy, {\it {Neutralino signatures of the
  singlet extended MSSM}},  {\em Phys. Lett.} {\bf B644} (2007) 361--369,
  [\href{http://xxx.lanl.gov/abs/hep-ph/0609068}{{\tt hep-ph/0609068}}].

\bibitem{Barger:2006sk}
V.~Barger, P.~Langacker, and G.~Shaughnessy, {\it {Collider Signatures of
  Singlet Extended Higgs Sectors}},  {\em Phys. Rev.} {\bf D75} (2007) 055013,
  [\href{http://xxx.lanl.gov/abs/hep-ph/0611239}{{\tt hep-ph/0611239}}].

\bibitem{Barger:2007nv}
V.~Barger, P.~Langacker, I.~Lewis, M.~McCaskey, G.~Shaughnessy, and B.~Yencho,
  {\it {Recoil Detection of the Lightest Neutralino in MSSM Singlet
  Extensions}},  {\em Phys. Rev.} {\bf D75} (2007) 115002,
  [\href{http://xxx.lanl.gov/abs/hep-ph/0702036}{{\tt hep-ph/0702036}}].

\bibitem{Balazs:2007pf}
C.~Balazs, M.~Carena, A.~Freitas, and C.~E.~M. Wagner, {\it {Phenomenology of
  the nMSSM from colliders to cosmology}},  {\em JHEP} {\bf 06} (2007) 066,
  [\href{http://xxx.lanl.gov/abs/0705.0431}{{\tt 0705.0431}}].

\bibitem{Ham:2007wu}
S.~W. Ham, J.~O. Im, and S.~K. OH, {\it {Electroweak phase transition in the
  MNMSSM with explicit CP violation}},
  \href{http://xxx.lanl.gov/abs/0707.4543}{{\tt 0707.4543}}.

\bibitem{Huber:2007vva}
S.~J. Huber and T.~Konstandin, {\it {Production of gravitational waves in the
  nMSSM}},  {\em JCAP} {\bf 0805} (2008) 017,
  [\href{http://xxx.lanl.gov/abs/0709.2091}{{\tt 0709.2091}}].

\bibitem{Chun:2008pg}
E.~J. Chun and P.~Roy, {\it {Dirac Leptogenesis in extended nMSSM}},  {\em
  JHEP} {\bf 06} (2008) 089, [\href{http://xxx.lanl.gov/abs/0803.1720}{{\tt
  0803.1720}}].

\bibitem{Ham:2008cg}
S.~W. Ham, J.~O. Im, and S.~K. OH, {\it {Neutral Higgs bosons in the MNMSSM
  with explicit CP violation}},  {\em Eur. Phys. J.} {\bf C58} (2008) 579--590,
  [\href{http://xxx.lanl.gov/abs/0805.1115}{{\tt 0805.1115}}].

\bibitem{Cao:2009ad}
J.~Cao, H.~E. Logan, and J.~M. Yang, {\it {Experimental constraints on nMSSM
  and implications on its phenomenology}},  {\em Phys. Rev.} {\bf D79} (2009)
  091701, [\href{http://xxx.lanl.gov/abs/0901.1437}{{\tt 0901.1437}}].

\bibitem{Cao:2010fi}
J.~Cao, K.-i. Hikasa, W.~Wang, J.~M. Yang, and L.-X. Yu, {\it {SUSY dark matter
  in light of CDMS II results: a comparative study for different models}},
  {\em JHEP} {\bf 07} (2010) 044,
  [\href{http://xxx.lanl.gov/abs/1005.0761}{{\tt 1005.0761}}].

\bibitem{Ishikawa:2014owa}
K.~Ishikawa, T.~Kitahara, and M.~Takimoto, {\it {Singlino Resonant Dark Matter
  and 125 GeV Higgs Boson in High-Scale Supersymmetry}},  {\em Phys. Rev.
  Lett.} {\bf 113} (2014), no.~13 131801,
  [\href{http://xxx.lanl.gov/abs/1405.7371}{{\tt 1405.7371}}].

\bibitem{Hesselbach:2007te}
S.~Hesselbach, D.~J. Miller, G.~Moortgat-Pick, R.~Nevzorov, and M.~Trusov, {\it
  {Theoretical upper bound on the mass of the LSP in the MNSSM}},  {\em Phys.
  Lett.} {\bf B662} (2008) 199--207,
  [\href{http://xxx.lanl.gov/abs/0712.2001}{{\tt 0712.2001}}].

\bibitem{Ellwanger:2005dv}
U.~Ellwanger and C.~Hugonie, {\it {NMHDECAY 2.0: An Updated program for
  sparticle masses, Higgs masses, couplings and decay widths in the NMSSM}},
  {\em Comput.Phys.Commun.} {\bf 175} (2006) 290--303,
  [\href{http://xxx.lanl.gov/abs/hep-ph/0508022}{{\tt hep-ph/0508022}}].

\bibitem{Ellwanger:2006rn}
U.~Ellwanger and C.~Hugonie, {\it {NMSPEC: A Fortran code for the sparticle and
  Higgs masses in the NMSSM with GUT scale boundary conditions}},  {\em
  Comput.Phys.Commun.} {\bf 177} (2007) 399--407,
  [\href{http://xxx.lanl.gov/abs/hep-ph/0612134}{{\tt hep-ph/0612134}}].

\bibitem{Kraml:2014sna}
S.~Kraml, S.~Kulkarni, U.~Laa, A.~Lessa, V.~Magerl, {\em et.~al.}, {\it
  {SModelS v1.0: a short user guide}},
  \href{http://xxx.lanl.gov/abs/1412.1745}{{\tt 1412.1745}}.

\bibitem{Kraml:2013mwa}
S.~Kraml, S.~Kulkarni, U.~Laa, A.~Lessa, W.~Magerl, {\em et.~al.}, {\it
  {SModelS: a tool for interpreting simplified-model results from the LHC and
  its application to supersymmetry}},  {\em Eur.Phys.J.} {\bf C74} (2014) 2868,
  [\href{http://xxx.lanl.gov/abs/1312.4175}{{\tt 1312.4175}}].

\bibitem{Conte:2012fm}
E.~Conte, B.~Fuks, and G.~Serret, {\it {MadAnalysis 5, A User-Friendly
  Framework for Collider Phenomenology}},  {\em Comput.Phys.Commun.} {\bf 184}
  (2013) 222--256, [\href{http://xxx.lanl.gov/abs/1206.1599}{{\tt 1206.1599}}].

\bibitem{Conte:2014zja}
E.~Conte, B.~Dumont, B.~Fuks, and C.~Wymant, {\it {Designing and recasting LHC
  analyses with MadAnalysis 5}},  {\em Eur.Phys.J.} {\bf C74} (2014), no.~10
  3103, [\href{http://xxx.lanl.gov/abs/1405.3982}{{\tt 1405.3982}}].

\bibitem{Dumont:2014tja}
B.~Dumont, B.~Fuks, S.~Kraml, S.~Bein, G.~Chalons, {\em et.~al.}, {\it {Toward
  a public analysis database for LHC new physics searches using MADANALYSIS
  5}},  {\em Eur.Phys.J.} {\bf C75} (2015), no.~2 56,
  [\href{http://xxx.lanl.gov/abs/1407.3278}{{\tt 1407.3278}}].

\bibitem{Belanger:2006is}
G.~Belanger, F.~Boudjema, A.~Pukhov, and A.~Semenov, {\it {MicrOMEGAs 2.0: A
  Program to calculate the relic density of dark matter in a generic model}},
  {\em Comput.Phys.Commun.} {\bf 176} (2007) 367--382,
  [\href{http://xxx.lanl.gov/abs/hep-ph/0607059}{{\tt hep-ph/0607059}}].

\bibitem{Belanger:2008sj}
G.~Belanger, F.~Boudjema, A.~Pukhov, and A.~Semenov, {\it {Dark matter direct
  detection rate in a generic model with micrOMEGAs 2.2}},  {\em
  Comput.Phys.Commun.} {\bf 180} (2009) 747--767,
  [\href{http://xxx.lanl.gov/abs/0803.2360}{{\tt 0803.2360}}].

\bibitem{Belanger:2014vza}
G.~Belanger, F.~Boudjema, A.~Pukhov, and A.~Semenov, {\it {micrOMEGAs4.1: two
  dark matter candidates}},  {\em Comput.Phys.Commun.} {\bf 192} (2015)
  322--329, [\href{http://xxx.lanl.gov/abs/1407.6129}{{\tt 1407.6129}}].

\bibitem{Chalons:2011ia}
G.~Chalons and A.~Semenov, {\it {Loop-induced photon spectral lines from
  neutralino annihilation in the NMSSM}},  {\em JHEP} {\bf 12} (2011) 055,
  [\href{http://xxx.lanl.gov/abs/1110.2064}{{\tt 1110.2064}}].

\bibitem{Das:2012ys}
D.~Das, U.~Ellwanger, and P.~Mitropoulos, {\it {A 130 GeV photon line from dark
  matter annihilation in the NMSSM}},  {\em JCAP} {\bf 1208} (2012) 003,
  [\href{http://xxx.lanl.gov/abs/1206.2639}{{\tt 1206.2639}}].

\bibitem{Chalons:2012xf}
G.~Chalons, M.~J. Dolan, and C.~McCabe, {\it {Neutralino dark matter and the
  Fermi gamma-ray lines}},  {\em JCAP} {\bf 1302} (2013) 016,
  [\href{http://xxx.lanl.gov/abs/1211.5154}{{\tt 1211.5154}}].

\bibitem{Giudice:1988yz}
G.~F. Giudice and A.~Masiero, {\it {A Natural Solution to the mu Problem in
  Supergravity Theories}},  {\em Phys. Lett.} {\bf B206} (1988) 480--484.

\bibitem{Vilenkin:1984ib}
A.~Vilenkin, {\it {Cosmic Strings and Domain Walls}},  {\em Phys. Rept.} {\bf
  121} (1985) 263--315.

\bibitem{Abel:1995wk}
S.~A. Abel, S.~Sarkar, and P.~L. White, {\it {On the cosmological domain wall
  problem for the minimally extended supersymmetric standard model}},  {\em
  Nucl. Phys.} {\bf B454} (1995) 663--684,
  [\href{http://xxx.lanl.gov/abs/hep-ph/9506359}{{\tt hep-ph/9506359}}].

\bibitem{Abel:1995uc}
S.~A. Abel and P.~L. White, {\it {Baryogenesis from domain walls in the
  next-to-minimal supersymmetric standard model}},  {\em Phys. Rev.} {\bf D52}
  (1995) 4371--4379, [\href{http://xxx.lanl.gov/abs/hep-ph/9505241}{{\tt
  hep-ph/9505241}}].

\bibitem{Nilles:1982mp}
H.~P. Nilles, M.~Srednicki, and D.~Wyler, {\it {Constraints on the Stability of
  Mass Hierarchies in Supergravity}},  {\em Phys. Lett.} {\bf B124} (1983) 337.

\bibitem{Lahanas:1982bk}
A.~B. Lahanas, {\it {Light Singlet, Gauge Hierarchy and Supergravity}},  {\em
  Phys. Lett.} {\bf B124} (1983) 341.

\bibitem{Ellwanger:1983mg}
U.~Ellwanger, {\it {Nonrenormalizable interactions from supergravity, quantum
  corrections and effective low-energy theories}},  {\em Phys. Lett.} {\bf
  B133} (1983) 187--191.

\bibitem{Nilles:1997me}
H.~P. Nilles and N.~Polonsky, {\it {Gravitational divergences as a mediator of
  supersymmetry breaking}},  {\em Phys. Lett.} {\bf B412} (1997) 69--76,
  [\href{http://xxx.lanl.gov/abs/hep-ph/9707249}{{\tt hep-ph/9707249}}].

\bibitem{Bagger:1993ji}
J.~Bagger and E.~Poppitz, {\it {Destabilizing divergences in supergravity
  coupled supersymmetric theories}},  {\em Phys. Rev. Lett.} {\bf 71} (1993)
  2380--2382, [\href{http://xxx.lanl.gov/abs/hep-ph/9307317}{{\tt
  hep-ph/9307317}}].

\bibitem{Jain:1994tk}
V.~Jain, {\it {On destabilizing divergencies in supergravity models}},  {\em
  Phys. Lett.} {\bf B351} (1995) 481--486,
  [\href{http://xxx.lanl.gov/abs/hep-ph/9407382}{{\tt hep-ph/9407382}}].

\bibitem{Bagger:1995ay}
J.~Bagger, E.~Poppitz, and L.~Randall, {\it {Destabilizing divergences in
  supergravity theories at two loops}},  {\em Nucl. Phys.} {\bf B455} (1995)
  59--82, [\href{http://xxx.lanl.gov/abs/hep-ph/9505244}{{\tt
  hep-ph/9505244}}].

\bibitem{Panagiotakopoulos:1998yw}
C.~Panagiotakopoulos and K.~Tamvakis, {\it {Stabilized NMSSM without domain
  walls}},  {\em Phys. Lett.} {\bf B446} (1999) 224--227,
  [\href{http://xxx.lanl.gov/abs/hep-ph/9809475}{{\tt hep-ph/9809475}}].

\bibitem{Degrassi:2009yq}
G.~Degrassi and P.~Slavich, {\it {On the radiative corrections to the neutral
  Higgs boson masses in the NMSSM}},  {\em Nucl. Phys.} {\bf B825} (2010)
  119--150, [\href{http://xxx.lanl.gov/abs/0907.4682}{{\tt 0907.4682}}].

\bibitem{Staub:2015aea}
F.~Staub, P.~Athron, U.~Ellwanger, R.~Grober, M.~Muhlleitner, P.~Slavich, and
  A.~Voigt, {\it {Higgs mass predictions of public NMSSM spectrum generators}},
   \href{http://xxx.lanl.gov/abs/1507.05093}{{\tt 1507.05093}}.

\bibitem{Belanger:2013xza}
G.~Belanger, B.~Dumont, U.~Ellwanger, J.~Gunion, and S.~Kraml, {\it {Global fit
  to Higgs signal strengths and couplings and implications for extended Higgs
  sectors}},  {\em Phys.Rev.} {\bf D88} (2013) 075008,
  [\href{http://xxx.lanl.gov/abs/1306.2941}{{\tt 1306.2941}}].

\bibitem{Aad:2015agg}
{\bf ATLAS} Collaboration, G.~Aad {\em et.~al.}, {\it {Search for a high-mass
  Higgs boson decaying to a $W$ boson pair in $pp$ collisions at $\sqrt{s} = 8$
  TeV with the ATLAS detector}},
  \href{http://xxx.lanl.gov/abs/1509.00389}{{\tt 1509.00389}}.

\bibitem{Adam:2015rua}
{\bf Planck} Collaboration, R.~Adam {\em et.~al.}, {\it {Planck 2015 results.
  I. Overview of products and scientific results}},
  \href{http://xxx.lanl.gov/abs/1502.01582}{{\tt 1502.01582}}.

\bibitem{Baro:2007em}
N.~Baro, F.~Boudjema, and A.~Semenov, {\it {Full one-loop corrections to the
  relic density in the MSSM: A Few examples}},  {\em Phys.Lett.} {\bf B660}
  (2008) 550--560, [\href{http://xxx.lanl.gov/abs/0710.1821}{{\tt 0710.1821}}].

\bibitem{Akerib:2013tjd}
{\bf LUX} Collaboration, D.~Akerib {\em et.~al.}, {\it {First results from the
  LUX dark matter experiment at the Sanford Underground Research Facility}},
  {\em Phys.Rev.Lett.} {\bf 112} (2014) 091303,
  [\href{http://xxx.lanl.gov/abs/1310.8214}{{\tt 1310.8214}}].

\bibitem{Allanach:2008qq}
B.~Allanach, C.~Balazs, G.~Belanger, M.~Bernhardt, F.~Boudjema, {\em et.~al.},
  {\it {SUSY Les Houches Accord 2}},  {\em Comput.Phys.Commun.} {\bf 180}
  (2009) 8--25, [\href{http://xxx.lanl.gov/abs/0801.0045}{{\tt 0801.0045}}].

\bibitem{Beenakker:1996ch}
W.~Beenakker, R.~Hopker, M.~Spira, and P.~Zerwas, {\it {Squark and gluino
  production at hadron colliders}},  {\em Nucl.Phys.} {\bf B492} (1997)
  51--103, [\href{http://xxx.lanl.gov/abs/hep-ph/9610490}{{\tt
  hep-ph/9610490}}].

\bibitem{Beenakker:1997ut}
W.~Beenakker, M.~Kramer, T.~Plehn, M.~Spira, and P.~Zerwas, {\it {Stop
  production at hadron colliders}},  {\em Nucl.Phys.} {\bf B515} (1998) 3--14,
  [\href{http://xxx.lanl.gov/abs/hep-ph/9710451}{{\tt hep-ph/9710451}}].

\bibitem{Kulesza:2008jb}
A.~Kulesza and L.~Motyka, {\it {Threshold resummation for squark-antisquark and
  gluino-pair production at the LHC}},  {\em Phys.Rev.Lett.} {\bf 102} (2009)
  111802, [\href{http://xxx.lanl.gov/abs/0807.2405}{{\tt 0807.2405}}].

\bibitem{Kulesza:2009kq}
A.~Kulesza and L.~Motyka, {\it {Soft gluon resummation for the production of
  gluino-gluino and squark-antisquark pairs at the LHC}},  {\em Phys.Rev.} {\bf
  D80} (2009) 095004, [\href{http://xxx.lanl.gov/abs/0905.4749}{{\tt
  0905.4749}}].

\bibitem{Beenakker:2009ha}
W.~Beenakker, S.~Brensing, M.~Kramer, A.~Kulesza, E.~Laenen, {\em et.~al.},
  {\it {Soft-gluon resummation for squark and gluino hadroproduction}},  {\em
  JHEP} {\bf 0912} (2009) 041, [\href{http://xxx.lanl.gov/abs/0909.4418}{{\tt
  0909.4418}}].

\bibitem{Beenakker:2010nq}
W.~Beenakker, S.~Brensing, M.~Kramer, A.~Kulesza, E.~Laenen, {\em et.~al.},
  {\it {Supersymmetric top and bottom squark production at hadron colliders}},
  {\em JHEP} {\bf 1008} (2010) 098,
  [\href{http://xxx.lanl.gov/abs/1006.4771}{{\tt 1006.4771}}].

\bibitem{Beenakker:2011fu}
W.~Beenakker, S.~Brensing, M.~Kramer, A.~Kulesza, E.~Laenen, {\em et.~al.},
  {\it {Squark and Gluino Hadroproduction}},  {\em Int.J.Mod.Phys.} {\bf A26}
  (2011) 2637--2664, [\href{http://xxx.lanl.gov/abs/1105.1110}{{\tt
  1105.1110}}].

\bibitem{Aad:2014vma}
{\bf ATLAS} Collaboration, G.~Aad {\em et.~al.}, {\it {Search for direct
  production of charginos, neutralinos and sleptons in final states with two
  leptons and missing transverse momentum in $pp$ collisions at $\sqrt{s} =$ 8
  TeV with the ATLAS detector}},  {\em JHEP} {\bf 1405} (2014) 071,
  [\href{http://xxx.lanl.gov/abs/1403.5294}{{\tt 1403.5294}}].

\bibitem{Chatrchyan:2013xna}
{\bf CMS} Collaboration, S.~Chatrchyan {\em et.~al.}, {\it {Search for
  top-squark pair production in the single-lepton final state in pp collisions
  at $\sqrt{s}$ = 8 TeV}},  {\em Eur.Phys.J.} {\bf C73} (2013), no.~12 2677,
  [\href{http://xxx.lanl.gov/abs/1308.1586}{{\tt 1308.1586}}].

\bibitem{Aad:2014nua}
{\bf ATLAS} Collaboration, G.~Aad {\em et.~al.}, {\it {Search for direct
  production of charginos and neutralinos in events with three leptons and
  missing transverse momentum in $\sqrt{s} =$ 8TeV $pp$ collisions with the
  ATLAS detector}},  {\em JHEP} {\bf 1404} (2014) 169,
  [\href{http://xxx.lanl.gov/abs/1402.7029}{{\tt 1402.7029}}].

\bibitem{MA5:CMS-SUS-13-016}
D.~Sengupta and S.~Kulkarni, {\it {MadAnalysis 5 implementation of
  CMS-SUS-13-016}}, .

\bibitem{CMS:2013ija}
{\bf CMS} Collaboration, C.~Collaboration, {\it {Search for supersymmetry in pp
  collisions at sqrt(s) = 8 Tev in events with two opposite sign leptons, large
  number of jets, b-tagged jets, and large missing transverse energy.}}, .

\bibitem{Alwall:2014hca}
J.~Alwall, R.~Frederix, S.~Frixione, V.~Hirschi, F.~Maltoni, {\em et.~al.},
  {\it {The automated computation of tree-level and next-to-leading order
  differential cross sections, and their matching to parton shower
  simulations}},  {\em JHEP} {\bf 1407} (2014) 079,
  [\href{http://xxx.lanl.gov/abs/1405.0301}{{\tt 1405.0301}}].

\bibitem{Sjostrand:2006za}
T.~Sjostrand, S.~Mrenna, and P.~Z. Skands, {\it {PYTHIA 6.4 Physics and
  Manual}},  {\em JHEP} {\bf 0605} (2006) 026,
  [\href{http://xxx.lanl.gov/abs/hep-ph/0603175}{{\tt hep-ph/0603175}}].

\bibitem{deFavereau:2013fsa}
{\bf DELPHES 3} Collaboration, J.~de~Favereau {\em et.~al.}, {\it {DELPHES 3, A
  modular framework for fast simulation of a generic collider experiment}},
  {\em JHEP} {\bf 1402} (2014) 057,
  [\href{http://xxx.lanl.gov/abs/1307.6346}{{\tt 1307.6346}}].

\bibitem{Cacciari:2011ma}
M.~Cacciari, G.~P. Salam, and G.~Soyez, {\it {FastJet User Manual}},  {\em Eur.
  Phys. J.} {\bf C72} (2012) 1896,
  [\href{http://xxx.lanl.gov/abs/1111.6097}{{\tt 1111.6097}}].

\bibitem{Cacciari:2008gp}
M.~Cacciari, G.~P. Salam, and G.~Soyez, {\it {The Anti-k(t) jet clustering
  algorithm}},  {\em JHEP} {\bf 04} (2008) 063,
  [\href{http://xxx.lanl.gov/abs/0802.1189}{{\tt 0802.1189}}].

\bibitem{Ackermann:2015zua}
{\bf Fermi-LAT} Collaboration, M.~Ackermann {\em et.~al.}, {\it {Searching for
  Dark Matter Annihilation from Milky Way Dwarf Spheroidal Galaxies with Six
  Years of Fermi-LAT Data}},  \href{http://xxx.lanl.gov/abs/1503.02641}{{\tt
  1503.02641}}.

\bibitem{Bi:2009uj}
X.-J. Bi, X.-G. He, and Q.~Yuan, {\it {Parameters in a class of leptophilic
  models from PAMELA, ATIC and FERMI}},  {\em Phys. Lett.} {\bf B678} (2009)
  168--173, [\href{http://xxx.lanl.gov/abs/0903.0122}{{\tt 0903.0122}}].

\bibitem{Cirelli:2013hv}
M.~Cirelli and G.~Giesen, {\it {Antiprotons from Dark Matter: Current
  constraints and future sensitivities}},  {\em JCAP} {\bf 1304} (2013) 015,
  [\href{http://xxx.lanl.gov/abs/1301.7079}{{\tt 1301.7079}}].

\bibitem{Aad:2015oqa}
{\bf ATLAS} Collaboration, G.~Aad {\em et.~al.}, {\it {Search for Higgs bosons
  decaying to $aa$ in the $\mu\mu\tau\tau$ final state in $pp$ collisions at
  $\sqrt{s} = $ 8 TeV with the ATLAS experiment}},
  \href{http://xxx.lanl.gov/abs/1505.01609}{{\tt 1505.01609}}.

\bibitem{Khachatryan:2015wka}
{\bf CMS} Collaboration, V.~Khachatryan {\em et.~al.}, {\it {A search for pair
  production of new light bosons decaying into muons}},
  \href{http://xxx.lanl.gov/abs/1506.00424}{{\tt 1506.00424}}.

\bibitem{Bomark:2014gya}
N.-E. Bomark, S.~Moretti, S.~Munir, and L.~Roszkowski, {\it {A light NMSSM
  pseudoscalar Higgs boson at the LHC redux}},  {\em JHEP} {\bf 1502} (2015)
  044, [\href{http://xxx.lanl.gov/abs/1409.8393}{{\tt 1409.8393}}].

\bibitem{Aprile:2012zx}
{\bf XENON1T} Collaboration, E.~Aprile, {\it {The XENON1T Dark Matter Search
  Experiment}},  {\em Springer Proc.Phys.} {\bf C12-02-22} (2013) 93--96,
  [\href{http://xxx.lanl.gov/abs/1206.6288}{{\tt 1206.6288}}].

\bibitem{Ruppin:2014bra}
F.~Ruppin, J.~Billard, E.~Figueroa-Feliciano, and L.~Strigari, {\it
  {Complementarity of dark matter detectors in light of the neutrino
  background}},  {\em Phys.Rev.} {\bf D90} (2014), no.~8 083510,
  [\href{http://xxx.lanl.gov/abs/1408.3581}{{\tt 1408.3581}}].

\bibitem{Harlander:2012pb}
R.~V. Harlander, S.~Liebler, and H.~Mantler, {\it {SusHi: A program for the
  calculation of Higgs production in gluon fusion and bottom-quark annihilation
  in the Standard Model and the MSSM}},  {\em Comput. Phys. Commun.} {\bf 184}
  (2013) 1605--1617, [\href{http://xxx.lanl.gov/abs/1212.3249}{{\tt
  1212.3249}}].

\bibitem{Liebler:2015bka}
S.~Liebler, {\it {Neutral Higgs production at proton colliders in the
  CP-conserving NMSSM}},  {\em Eur. Phys. J.} {\bf C75} (2015), no.~5 210,
  [\href{http://xxx.lanl.gov/abs/1502.07972}{{\tt 1502.07972}}].

\bibitem{Khachatryan:2014qwa}
{\bf CMS} Collaboration, V.~Khachatryan {\em et.~al.}, {\it {Searches for
  electroweak production of charginos, neutralinos, and sleptons decaying to
  leptons and W, Z, and Higgs bosons in pp collisions at 8 TeV}},  {\em
  Eur.Phys.J.} {\bf C74} (2014), no.~9 3036,
  [\href{http://xxx.lanl.gov/abs/1405.7570}{{\tt 1405.7570}}].

\bibitem{Khachatryan:2015lba}
{\bf CMS} Collaboration, V.~Khachatryan {\em et.~al.}, {\it {Search for a
  pseudoscalar boson decaying into a Z boson and the 125 GeV Higgs boson in
  llbb final states}},  \href{http://xxx.lanl.gov/abs/1504.04710}{{\tt
  1504.04710}}.

\bibitem{Aad:2015wra}
{\bf ATLAS} Collaboration, G.~Aad {\em et.~al.}, {\it {Search for a CP-odd
  Higgs boson decaying to Zh in pp collisions at $\sqrt{s} = 8$ TeV with the
  ATLAS detector}},  {\em Phys.Lett.} {\bf B744} (2015) 163--183,
  [\href{http://xxx.lanl.gov/abs/1502.04478}{{\tt 1502.04478}}].

\bibitem{Khachatryan:2014jya}
{\bf CMS} Collaboration, V.~Khachatryan {\em et.~al.}, {\it {Searches for heavy
  Higgs bosons in two-Higgs-doublet models and for $t \to ch$ decay using
  multilepton and diphoton final states in $pp$ collisions at 8 TeV}},  {\em
  Phys. Rev.} {\bf D90} (2014) 112013,
  [\href{http://xxx.lanl.gov/abs/1410.2751}{{\tt 1410.2751}}].

\bibitem{Heinemeyer:2013tqa}
{\bf LHC Higgs Cross Section Working Group} Collaboration, S.~Heinemeyer {\em
  et.~al.}, {\it {Handbook of LHC Higgs Cross Sections: 3. Higgs Properties}},
  \href{http://xxx.lanl.gov/abs/1307.1347}{{\tt 1307.1347}}.

\bibitem{Barger:2006hm}
V.~Barger, T.~Han, and D.~G.~E. Walker, {\it {Top Quark Pairs at High Invariant
  Mass: A Model-Independent Discriminator of New Physics at the LHC}},  {\em
  Phys. Rev. Lett.} {\bf 100} (2008) 031801,
  [\href{http://xxx.lanl.gov/abs/hep-ph/0612016}{{\tt hep-ph/0612016}}].

\bibitem{Djouadi:2015jea}
A.~Djouadi, L.~Maiani, A.~Polosa, J.~Quevillon, and V.~Riquer, {\it {Fully
  covering the MSSM Higgs sector at the LHC}},  {\em JHEP} {\bf 06} (2015) 168,
  [\href{http://xxx.lanl.gov/abs/1502.05653}{{\tt 1502.05653}}].

\bibitem{Belanger:2015vwa}
G.~Belanger, D.~Ghosh, R.~Godbole, and S.~Kulkarni, {\it {Light stop in the
  MSSM after LHC Run 1}},  \href{http://xxx.lanl.gov/abs/1506.00665}{{\tt
  1506.00665}}.

\bibitem{Aad:2014kra}
{\bf ATLAS} Collaboration, G.~Aad {\em et.~al.}, {\it {Search for top squark
  pair production in final states with one isolated lepton, jets, and missing
  transverse momentum in $\sqrt s =$8 TeV $pp$ collisions with the ATLAS
  detector}},  {\em JHEP} {\bf 1411} (2014) 118,
  [\href{http://xxx.lanl.gov/abs/1407.0583}{{\tt 1407.0583}}].

\bibitem{MA5:CMS-SUS-13-011}
B.~Dumont, B.~Fuks, and C.~Wymant, {\it {MadAnalysis 5 implementation of
  CMS-SUS-13-011: search for stops in the single lepton final state at 8 TeV}},
  .

\bibitem{Aad:2014yka}
{\bf ATLAS} Collaboration, G.~Aad {\em et.~al.}, {\it {Search for the direct
  production of charginos, neutralinos and staus in final states with at least
  two hadronically decaying taus and missing transverse momentum in $pp$
  collisions at $\sqrt{s}$ = 8 TeV with the ATLAS detector}},  {\em JHEP} {\bf
  1410} (2014) 96, [\href{http://xxx.lanl.gov/abs/1407.0350}{{\tt 1407.0350}}].

\end{thebibliography}\endgroup

\end{document}